# An Exploration of IoT Platform Development

Dr. Mahdi Fahmideh[1], University of Wollongong, Australia
T: +61 2 87636496

Professor Didar Zowghi[2], University of Technology Sydney, Australia
T: +61 2 95141860

**Abstract.** IoT (Internet of Things) platforms are key enablers for smart city initiatives, targeting the improvement of citizens' quality of life and economic growth. As IoT platforms are dynamic, proactive, and heterogeneous socio-technical artefacts, systematic approaches are required for their development. Limited surveys have exclusively explored how IoT platforms are developed and maintained from the perspective of information system development process lifecycle. In this paper, we present a detailed analysis of 63 approaches. This is accomplished by proposing an evaluation framework as a cornerstone to highlight the characteristics, strengths, and weaknesses of these approaches. The survey results not only provide insights of empirical findings, recommendations, and mechanisms for the development of quality aware IoT platforms, but also identify important issues and gaps that need to be addressed.

# 1 Introduction

One of the key enablers of a smart city is the *IoT platforms* (du Plessis, 2018). An IoT platform is a set of technology-enabled entities including physical smart objects (e.g. sensors, actuators, cameras, smart tags, and tracking labels) as well as software services and systems that are connected and working together. An IoT platform, typically, collects and processes massive amount of data generated by smart city entities in a real-time fashion to improve city services to citizens (Williamson & Kennan, 2016), (Jin, Gubbi, Marusic, & Palaniswami, 2014). IoT platforms are a backbone for many smart cities such as those are in Europe (Caragliu, Del Bo, & Nijkamp, 2011), China (Liu & Peng, 2013), and United Arab Emirates (Janajreh, Su, & Alan, 2013).

An IoT platform may constitute millions of smart objects and software services that should operate in an orchestrated way to provide active sensing, and smart reasoning for citizens. As the development of such socio-technical artefacts is a complex and challenging process, the need for adopting systematic engineering approaches, i.e. engineering methodologies or information system development methods[3], to develop IoT platforms is pivotal (Zambonelli, 2016). Engineering approaches are the core of all well-engineered IT artefacts as they provide a means for applying practices, design decisions, and techniques for developing information systems (Avison & Fitzgerald, 2003). Analogically, it is evident that an IoT platform development is, after all, essentially a type of information system development (Savaglio, 2017), (Diaconescu & Wagner, 2014). Considering this analogy, adopting an engineering lifecycle perspective for managing the complexity of IoT platform development is acclaimed  as it takes precedence over an ad-hoc use of implementation techniques and technologies which are likely to deliver a vulnerable and poor quality platform (Savaglio, 2017). This has been acknowledged by earlier research suggesting IoT development should be conducted from the engineering lifecycle point of view (Wenge, Zhang, Dave, Chao, & Hao, 2014). According to the Gartner's report (Pettey, 2018):

> *"...developing and standardizing the process for building IoT solutions and then guiding the evolution and improvement of that process is key. This will help make the organization's creation of IoT solutions easier and more reliable because these initiatives will follow a process that incorporates the organization's experience and accrued best practices in IoT solution development."*

---

[1] Lecturer (Assistant Professor) of Information Technology in the school of computing and information technology, Faculty of Engineering and Information Science, University of Wollongong, Wollongong, Australia

[2] Professor of Software Engineering in the Faculty of Engineering and Information Technology at the University of Technology Sydney (UTS), Sydney, Australia
[3] As it will be explained in section 4.2, the term approach is used to refer method, methodology, platforms, and conceptual models.

Moreover, Fortino et. al., who designed an approach for a smart tourism IoT platform, state (Fortino, Guerrieri, Russo, & Savaglio, 2015):

> *"...to fully exploit the widely recognized smart objects' potential in analysing, designing and implementing IoT eco-systems, well-defined development methodologies are required."*

Nevertheless, the development processes for IoT has not yet been explored as the hype suggests. Practitioners may be arguably referred to traditional engineering lifecycles (e.g. SDLC) and software engineering practices to develop an IoT platform. However, as it will be discussed in Section 4, an IoT platform development endeavour is distinct from the traditional information system development in several ways. Software components, mobile applications, and backbone services, that are combined together to offer IoT services, are developed and maintained in a typical information systems project. On the other hand, hardware components of a platform should be able to communicate with other software components, which can be a complete project on its own and thus needs to be developed and maintained via a different lifecycle. Apart from technical challenges, an IoT platform development may involve multiple domains and thus a diversity of stakeholders and their requirements (Slama, Puhlmann, Morrish, & Bhatnagar, 2015). Bringing these different lifecycles together implies the need for engineering new engineering approaches or augmenting existing ones to incorporate and address the abovementioned issues in the course of an IoT platform development.

There is an on-going proliferation of approaches as reviewed in Section 4. They may often give too little or too many details at different levels of abstraction that makes hard perceiving and explaining the underlying mainstream of IoT development process. Limited existing surveys have exclusively devoted their effort to understand what certain aspects and requirements should be taken into account in an IoT development lifecycle or assessing the suitability of an existing approach for a specific smart city project. Hence, a review of the literature in this field, to identify research gaps in the current landscape and inform future research on this topic is timely. We aim to identify what is already known about the development process lifecycle of IoT platforms, synthesize the current research evidence, and propose an agenda for future studies. This aim is accommodated by means of an evaluation framework through which existing approaches are compared and contrasted. Accordingly, this paper attempts to answer the following research questions:

–RQ. What is the current state of existing approaches for developing IoT platforms with respect to the proposed evaluation framework introduced in Section 2?

RQ1. what is the application and type of these approaches?

RQ2. How is IoT platform development process lifecycle perceived in the literature?

RQ3. what roles are involved in the development of IoT platforms?

RQ4. what modelling activities and modelling languages are used during IoT platform development?

We answer these research questions and present the evaluation results of a representative set of existing IoT approaches using a proposed evaluation framework. As it will be discussed in Section 7, the focus, depth and coverage of our analysis and the survey coverage have not been provided by prior research. We, therefore, position this survey as the unprecedented reference point contributing to the literature on three aspects.

— This is the first research that sheds light on the typical development lifecycle of IoT platforms by providing a comprehensive review and synthesis of commonly occurring activities, quality factors, and recommendations enabling scholars and practitioners (e.g. platform providers), to understand challenges and employ techniques to tackle these challenges.

— The proposed evaluation framework comprises a set of features to categorise and examine IoT development approaches. The framework characterises different features for the incorporation into the development of an IoT platform which are helpful for platform providers to compare existing approaches or check if their own in-house approach is suitable to implement and maintain an IoT platform. In other words, the survey provides a knowledge base that can be helpful for platform providers to design a bespoke IoT development approach. A key feature of the framework is its abstraction and being independent of specific standards, technologies, and implementations.

— This survey extends the previous ones by adding the distinguishing development lifecycle point of view to the IoT literature. The identified gaps in this domain, open new research opportunities for researchers.

This article is organised as follows: Section 2 presents our evaluation framework designed for the purpose of this survey. Section 3 presents the systematic survey used to conduct this research. Section 4 discusses how the existing works address the different features of the framework and it delineates the recommendations to for effective IoT platform development. This is followed by a discussion on the survey findings and remaining challenges pointing to further research directions in Section 5. The research threats discussed in Section 6. Section 7 reviews the previously published surveys that are related to the one we present. Finally, this article concludes in Section 8.

## 2 Evaluation framework

We propose our evaluation framework leaning heavily towards assessing engineering lifecycle processes to let us classifying, analysing, and characterising existing IoT development approaches and thus answer to the research questions. The construction of the evaluation framework was conducted in three steps as described in the following subsections.

### 2.1 Step 1. Defining meta-features

We sought desirable features that are expected to be satisfied by an ideal evaluation framework. Such features, called *meta-features*, are used to evaluate other features. The definition of meta-features may depend on a domain context; however, having a list of them to be used during the compilation of a feature set to get a fair framework is essential. We used the following meta-features defined by (Karam & Casselman, 1993) and (Taromirad & Ramsin, 2008) during compiling the evaluation framework: (i) *simplicity*, i.e. a feature should be clear and easy to understand, (ii) *preciseness*, i.e. a feature should be detailed, unambiguous, and quantifiable to be usable by assessors, (iii) *minimum overlapping*, i.e. features should be distinct and have minimum dependency to each other, (iv) *soundness*, i.e. a feature should be related to and have semantic link to the problem domain, and (v) *generality*, i.e. a feature should be abstract and independent of specific standards, technologies, implementations, and other concrete details.

### 2.2 Step 2. Derivation of feature set

The compilation of the framework's features was inspired by the existing evaluation frameworks for system development approaches, but it was specialised for the context of IoT platform development. As mentioned in Section 1, an IoT platform development endeavour can be comparatively viewed as a particular class of software system development endeavour. To achieve a set of features which adhere to the meta-features (step 2.1), we began to identify a few overarching and workable dimensions as an overall frame to group the features. The development of the features was conducted in two steps as follows.

We reviewed the existing traditional evaluation frameworks such as (Karam & Casselman, 1993), (Ramsin & Paige, 2008), (Sturm & Shehory, 2004), and (Pressman, 2005) and synthesised the features suggested by these sources to derive a fair set of features that could be sufficiently abstract, application independent, and equally applicable in the context of an IoT platform development. The output of this step was a general evaluation framework that subsumes the features under the following four aspects that are elaborated throughout Section 4:

— *Context* characterises the application domain and geographical location for which an approach has been designed;
— *Lifecycle coverage* ascertains phases performing for an IoT platform development;
— *Roles* describe different human entities that are involved during a platform development;
— *Modelling* captures various models and representational languages used for data and work-products in a platform development process.

The feature of *lifecycle coverage* in the evaluation framework is inspired by the generic software development phases (Pressman, 2005) and it is broken down into *initialisation*, *analysis*, *design*, *implementation and test*, and *deployment*. We strived to extend this feature to more detailed features that were highly related and deemed important for an IoT platform development. This resulted in the derivation of the new features under the *lifecycle coverage* in the framework. These features that were deeply influenced by previous works such as (da Silva et al., 2013), (Al-Fuqaha, Guizani, Mohammadi, Aledhari, & Ayyash, 2015) and closest research domains to IoT, in particular, the feature set proposed in (Fahmideh, Daneshgar, Low, & Beydoun, 2016) are: *resource discovery*, *data management*, *monitoring*, *service composition*, and *event processing*.

It should be noted that the derivation of the feature set was conducted in a (i) top-down manner, i.e. reviewing introductory papers such as literature surveys highlighting key challenges of IoT development and (ii) bottom-up manner by examining different existing IoT platforms and development approaches reviewed in Section 4. Furthermore, the feature derivation has been iteratively influenced by the presented platforms in Section 4 in the sense that reviewing them led us to more in-depth understanding of the identified features and thus resulted in further refinements and extensions of the features. The iterations for refining the feature were continued until they got to sufficiently stabilised so that further iterations did not resulted in new features. We excluded the features that seem to be purely technical or platform dependent. For instance, we believed that the feature called *wireless sensor network management*, defined in (Kyriazopoulou, 2015), can be covered by our feature *resource discovery* in the proposed framework and thus it was removed from our feature set. Figure 1 shows the resultant evaluation framework that provides a high-level frame to compare and classify the existing IoT development approaches.

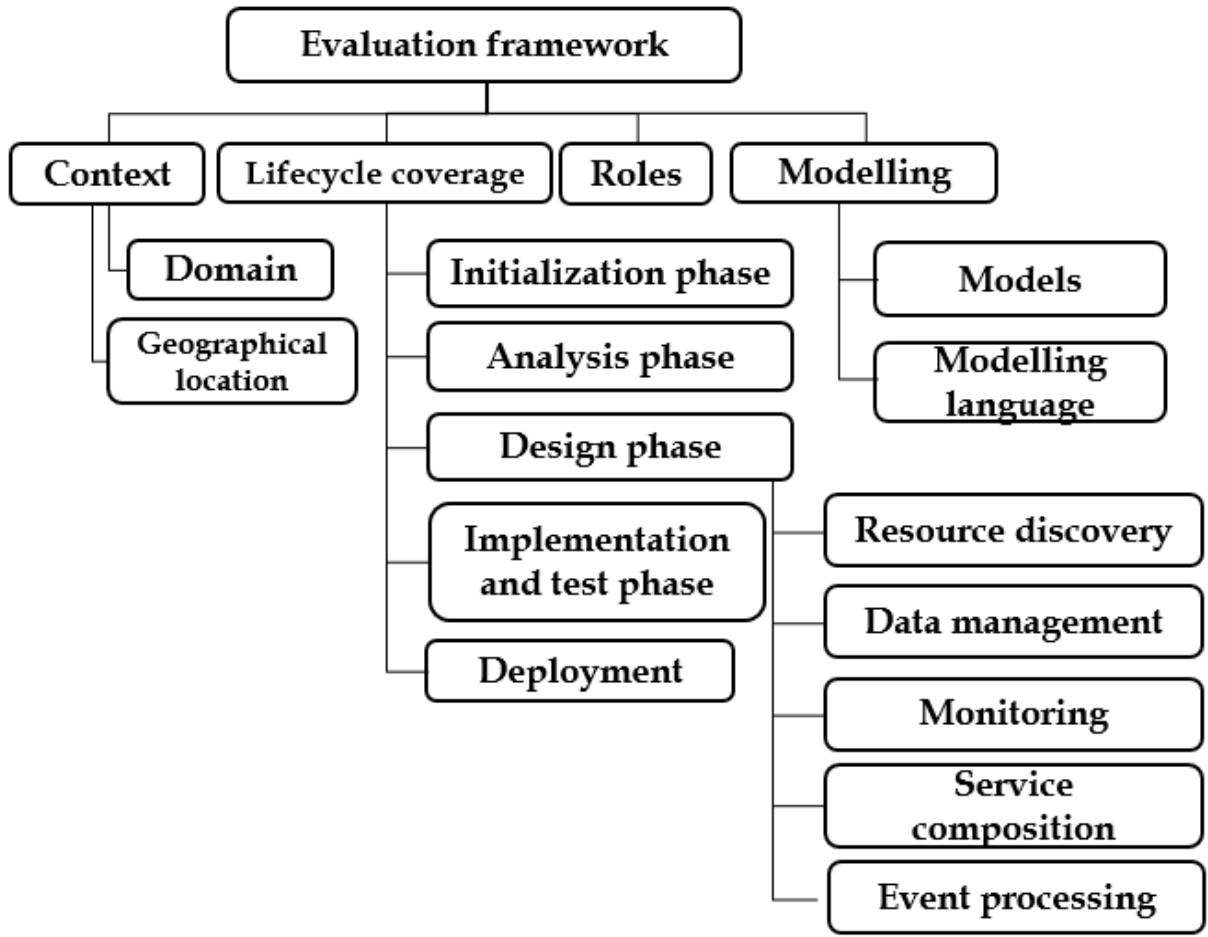

Figure 1. Evaluation framework for analysing IoT platform development approaches

## 2.3 Step 3. Evaluation of feature set

We obtained qualitative feedback from domain experts in the IoT field regarding the framework's adherence to the meta-features defined in Step 1. This gave us an opportunity to refine the framework. The following criteria were set to select a domain expert: (i) software developers/architects with real world experience in IoT platform development, or (ii) an academic with an extensive domain knowledge acknowledged by their publications in IoT related journals and conferences venues, and (iii) having good command of English language. Two volunteer domain experts, associated to our project which denoted as E1 and E2, accepted to independently assess the document of our framework based on the meta-features. The profile of reviewers who were both from Sydney, Australia, is as follows. The first reviewer was a university senior lecturer who published works in IoT related venues and had real-world experience in conducting enterprise architecture design. The second reviewer was a senior research consultant in IoT industry. The review process was taken between July and September 2018. A questionnaire form including questions related to the meta-features was given to the domain experts to read and provide their comments. Overall feedback was positive confirming that the feature set is sound and have an applicability to assess approaches. Some experts' comments were deemed out of the scope of this research though they were valuable for the further extension of the framework. For example, E1 suggested the creation of an online version of the framework as well as re-framing the framework's dimension regarding common architectural frameworks such as TOGAF. The framework was used to evaluate the existing IoT development approaches as discussed in Section 4.

# 3 Systematic survey

## 3.1 Overview

We conducted a systematic literature review (Figure 2); this followed the procedure for data collection and analysis described by (Kitchenham et al., 2009). First, scientific digital libraries were searched against the inclusion criteria and proposed search strings derived from the research questions. Next, research papers that met the inclusion criteria were selected. Third, data from selected papers were extracted and, finally, qualitative analysis was performed. The following sections discuss the search strategy and data synthesis.

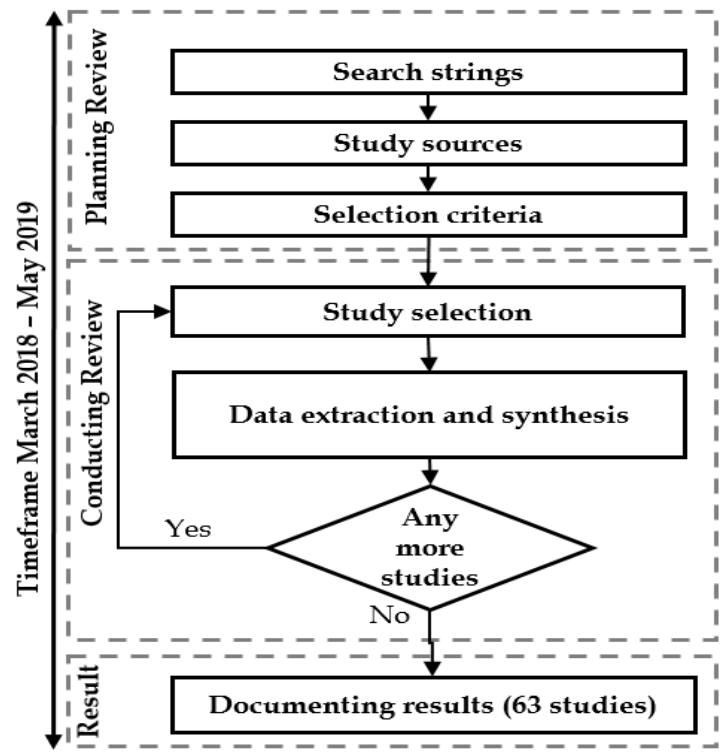

Figure 2. Systematic literature review conducted for this research

## 3.2 Planning review

### Search string

From an initial screening of the literature, we realized that existing works in literature do not necessarily use the same terminologies to refer to engineering lifecycle for implementing IoT platforms. In IoT literature, a platform development has been described in different abstraction levels, the granularity of concepts/layers, and combined with enabling technical platforms. This could lead to identifying too many studies or to miss some important ones. Hence, defining search strings was challenging. As a countermeasure, we continually refined the search strings to avoid missing any papers. Following guidelines by (Dieste & Padua, 2007), we first determined the main terms by decomposing the research questions. The main terms *IoT platform* and *approach* were extended with alternative synonyms. Both were combined to define a set of search queries using logical operations AND and OR and to be used against the title, abstract and keywords of the studies during the conducting review phase. The search queries shown in Table 1.

Table 1. Search strings

| Search Query (SQ) |
|---|
| **SQ1:** "IoT", OR "IoT platform" OR "Platform" OR "Smart city IOT" AND [SQ2] |
| **SQ2:** "Approach" OR "Method" OR "Methodology" OR "Information System Development Method" OR "Software Development Methodology" OR "Process" OR "Development Process" OR "Process |

Model" OR "Process Lifecycle" OR "Lifecycle" OR "Reference Model" OR "Framework" OR "Engineering Methodology" OR "Engineering Method"

## Study sources

The common scientific digital libraries IEEE Explore, ACM Digital Library, SpringerLink, ScienceDirect, Wiley InterScience, ISI Web of Knowledge, and Google Scholar were defined as sources for the literature search. These libraries maintain the majority of published studies in IoT and software engineering lifecycle approaches. We took into account articles published in the prestigious IS and SE journals. For this particular study, the proceedings of international conferences, symposiums, and workshops related to IoT platform development were used and proved to be a very useful source of information. We also took into account online non-academic literature, called multi-vocal literature (Ogawa & Malen, 1991), such as internet blogs, white papers, and trade journal articles, which could readily propose ideas surrounding IoT platform development.

## Selection criteria

While our survey focused on studies pertaining to the development process of IoT platforms, we selected identified studies that meet the following criteria:

—research questions/objective sufficiently described;
—findings and conclusions well explained;
—the development process or architecture design for IoT platforms explained including activities and mechanisms; and
—published between 2008 and May 2019.

As this survey focused on studies related to the development of IoT platforms, the following exclusion criteria were set for:

—studies in languages other than English;
—introductory papers to smart city development and IoT architecture; and

## 3.2 Conducting review

### Study selection

To avoid missing any related paper, conducting the literature review was followed by performing the reference snowballing technique (Wohlin, 2014) in the sense that studies cited in the references and related work section of the paper were feeding into the next run of the literature search. Moreover, the studies that cited the current study were identified. This phase conducted in several back-and-forths through refining the search strings and the literature search (Figure 2). The literature search was not performed merely based on automated search, but also included an extensive manual search. We also identified some IoT standards that have not been published in the academic literature through the manual search in the selected sources. The rationale to include this bunch of standards was to assort a representative mix of well-published industrial IoT platforms along with ones proposed by academia. We identified 125 papers. However, after removing duplications 63 papers were left. We read the abstract, introduction, and conclusion of each paper and evaluated it against the inclusion/exclusion criteria.

### data extraction and synthesis

We used the evaluation framework as a lens to extract key data from 63 identified studies. We imported the data into Excel sheets to capture the full details of the studies under review. The full text of each study was thoroughly read and corresponding text segments such as sentences, phrases, or paragraphs, which were relevant to a feature were extracted along with the reference to the study. Apart from that, data items pertaining to the research quality were extracted for further assessment as discussed in Section 4.1. We used the criteria defined in Critical Appraisal Skills Programme (Greenhalgh & Taylor, 1997) along with those suggested for conducting evidence-based software engineering (Kitchenham et al., 2002). Full demographic information of the studies including authors, title, acronym, publication channel, and source year is presented in Appendix A.

# 4 Results

Section 4.1 gives an overview of the research quality of the selected approaches. Section 4.2 summarises the analysis results using the evaluation framework. Sections 4.3, 4.4, 4.5, and 4.6 describe the analysis of the studies to answer the research questions using each feature of the proposed evaluation framework and thus to propose recommendations, lessons learned, and mechanisms for IoT platform development.

## 4.1 Research quality

We used eight criteria adopted from (Greenhalgh & Taylor, 1997) and (Kitchenham et al., 2002) to assess the quality of the research design used. The list of questions presented in Appendix B was used to grade the satisfaction of each criterion. The assessment results have been described based on five scales *completely addressed*, *considerably addressed*, *moderately addressed*, *slightly addressed*, and *not addressed*. Figure 3 shows the distribution of the criteria satisfaction by the approaches. According to this figure, the majority of the approaches have clearly stated a research aim. As far as the criterion research design is concerned, only 8 studies [S8], [S17], [S38], [S49], [S57], [S60], [SS62], and [S63], i.e. (13%), have provided either a full or considerable description of the research design to conduct their work. As many as 35 of 63 studies did not describe the research design at all. Additionally, an overall view of the scores in Figure 3 reveals that a large number of studies have not addressed the criteria data collection, data analysis, and reflexivity. To be more specific, 36 studies have not described the way data was collected and analysed to validate the proposed IoT development approach. From Figure 3 it can be observed that 49 studies, i.e. 84%, do not report how researchers have been involved with the environment in which the validation of a proposed approach conducted. We believe that the research design to propose approaches for IoT platform development has been viewed as a subsidiary task.

According to Figure 3, 70% of studies explicitly described their contributions to the literature. Studies [S15], [S60], and [S63] achieved the best three scores on the quality assessment whilst studies [S22], [S23], [S30], and [S61] received the lowest score in this review. For the criterion validation, the studies were classified according to the applied validation type as shown in Table 2. The majority of approaches applied the case study example (33), followed by techniques such as experience report (10), simulation (4), theoretical validation (2), and workshop (1). Of 63 studies, 13 (22%) did not present any information on the validation. Whilst including studies with a poor validation might be counted as a violation from recommended the systematic literature review procedure, we tended to cover the entire relevant literature as much as possible.

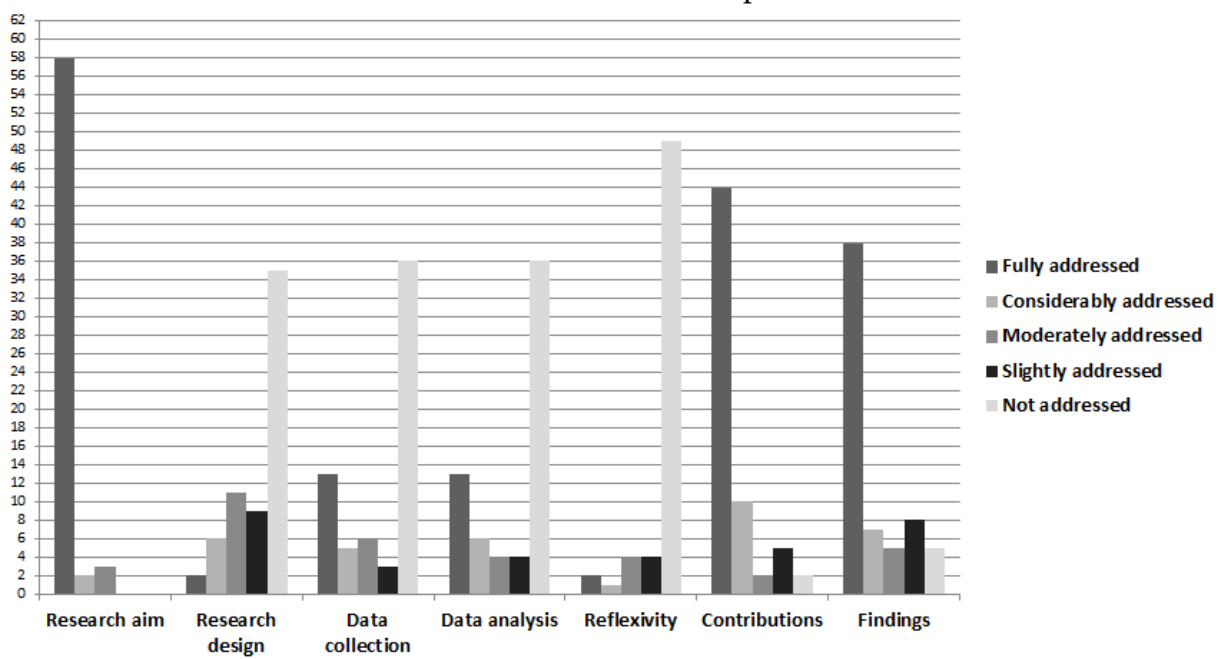

Figure 3. quality scores for the identified studies

Table 2. validation type used in existing approaches

| Study | Validation | Description | Number |
|---|---|---|---|
| [S1], [S2], [S5], [S7], [S9], [S14], [S15], [S18], [S20], [S21], [S23], [S24], [S25], [S26], [S27], [S28], [S37], [S38], [S39], [S42], [S44],[S47],[S49],[S51],[S52],[S53],[S54],[S55],[S56],[S57], [S61], [S62], [S63] | Case study | Case study has been used to investigate the approach within real-life or exemplar context. | 33 |
| [S6], [S8], [S12], [S33], [S36], [S41], [S45], [S48], [S50], [S59] | Experience report | The approach has been developed on the basis of gained experience in an industrial experience. | 10 |
| [S3], [S32], [S43], [S60] | Simulation | A mathematic simulation used to assess the approach correctness. | 4 |
| [S10], [S19] | Theoretical validation | The approach has been validated using a set of high-level criteria. | 2 |
| [S34] | Workshop | The approach presented to domain experts and it received feedback | 1 |
| [S4], [S11], [S13], [S16], [S17], [S22], [S29], [S30], [S31], [S35], [S40], [S46], [S58] | Not validated | The approach did not specify any applied validation | 13 |

## 4.2 Overview

Since we realized that the knowledge about the way of developing IoT platforms is spread out over the literature where each piece of work provides a different level of sophistication in developing IoT platforms varying from very high-level to technical level of details, we used the term "approach" as an overarching term to refer to any systematic way of developing IoT platforms. Hence, the term of approach includes any of these:

— *Conceptual model* is the most basic and high-level presentation of an IoT architecture, which is not dependent or bound to a specific domain or technology. An approach at this level gives platform providers an overall view of an IoT platform architecture components. RASCP [S1], MC-IoT [S7], EADIC [S17], SCRM [S19], TMN [S31], RAMI [S33], BSI [45] and standardisation efforts such as SCCM [S40] are some examples of this class.

— *Platforms* which may focus on either (i) software components including application programming interfaces (APIs), services, and tools in order to develop real IoT software applications (e.g. VITAL [S2], CiDAP [S3], Cisco [S6], IoT-ARM [S8], OpenIoT [S9], FIWARE [S12]) or (ii) hardware and infrastructure components to integrate heterogeneous and geographically dispersed smart objects via network and electronic protocols (e.g. Telco USN-Platform [S18], SPITFIRE [S20], and Padova [S24]).

— *Method (or engineering methodology)* define activities and guidance to implement an IoT platform. For instance, TSB [S36], BSI [S45], and ESPRESSO [S48] define delivering strategies to transform city services to future IoT based applications.

Table 3 presents the analysis results of the approaches with respect to the evaluation framework. These results have been in tables 3,4, and 5. Symbols √ and × in Table 3 indicate if a feature is/is not supported by an approach. This support means that the approach has provided mechanisms or techniques in order to address that feature. Note that, in this survey, we denote each studied approach using its abbreviation (Appendix A) along with a unique identifier starting with 'S'. These identifiers are used throughout the article.

Table 3. Evaluation of the existing approaches against the evaluation framework, note: - √: addressed, ×:not-addressed, C: Conceptual model, P:Platform, M:Method

| Study Id | Proposal acronym/name | Type | Initialise | Analyse | Resource discovery | Resource management | Data accumulation | Data cleaning | Data storing | Data processing | Query processing | Meta-data generation | Data visualisation | Monitoring | Service composition | Event processing | Implementation & test | Deployment | Roles | Modelling language | Models |
|---|---|---|---|---|---|---|---|---|---|---|---|---|---|---|---|---|---|---|---|---|---|
| [S1] | RASCP | C | × | × | √ | × | √ | √ | √ | √ | × | √ | √ | × | × | √ | √ | × | × | × | × |
| [S2] | VITAL | P | × | × | √ | × | √ | √ | √ | √ | × | √ | √ | √ | √ | √ | × | × | × | √ | √ |
| [S3] | CiDAP | P | × | × | × | √ | √ | √ | √ | √ | √ | × | √ | √ | × | × | × | × | × | × | √ |
| [S4] | Khan | C | × | × | √ | × | √ | √ | √ | √ | × | √ | × | × | × | × | × | × | × | × | × |
| [S5] | Gubbi | P | × | × | √ | × | √ | √ | × | × | × | × | × | × | × | × | × | × | × | × | × |
| [S6] | Cisco | P | × | × | × | × | √ | √ | × | × | × | √ | √ | √ | × | √ | × | × | × | × | × |
| [S7] | MC-IoT | C | × | × | × | × | × | × | × | × | × | × | × | × | × | × | × | × | × | × | × |
| [S8] | IoT-ARM | P | √ | √ | √ | × | × | × | × | × | × | × | × | × | √ | × | × | × | × | √ | √ |
| [S9] | OpenIoT | P | × | × | √ | × | √ | √ | √ | √ | √ | √ | √ | √ | × | × | × | × | × | √ | × |
| [S10] | Guth | C | × | × | × | × | × | × | × | × | × | × | × | × | × | × | × | × | × | × | × |
| [S11] | Ganchev | C | × | × | × | × | × | × | × | × | × | × | × | × | × | × | × | × | × | × | × |
| [S12] | FIWARE/OASC | P | × | × | × | × | × | × | √ | × | √ | × | × | × | √ | √ | √ | × | × | × | × |
| [S13] | Vilajosana | C | × | × | × | × | × | × | × | × | × | × | × | × | × | × | × | √ | × | × | √ |
| [S14] | Scallop4SC | P | × | × | × | × | × | × | √ | × | √ | × | × | × | × | × | × | × | × | × | × |
| [S15] | Khan | P | × | × | × | × | √ | √ | × | × | × | √ | × | × | √ | × | × | × | × | × | × |
| [S16] | Catherine | C | × | × | × | × | × | × | × | × | × | × | × | × | × | × | × | × | × | × | × |
| [S17] | EADIC | C | × | × | × | × | × | × | × | × | × | × | × | × | × | × | × | × | × | √ | √ |
| [S18] | Telco USN-Platform | P | × | × | √ | × | × | × | × | × | × | × | × | × | × | × | × | × | × | × | × |
| [S19] | SCRM | C | √ | × | × | × | × | × | × | × | × | × | × | × | × | × | × | × | × | × | × |
| [S20] | SPITFIRE | P | × | × | × | × | × | × | × | × | √ | × | × | × | × | × | × | × | × | √ | × |
| [S21] | Giang | P | × | × | × | × | × | × | × | × | × | × | × | × | × | × | × | × | × | × | √ |
| [S22] | Wenge | C | × | × | × | × | √ | × | √ | × | × | × | × | × | √ | × | √ | × | √ | × | × |
| [S23] | WSO2 | P | × | × | × | × | √ | × | × | × | × | × | × | × | × | √ | × | × | × | × | × |
| [S24] | Padova | P | × | × | × | × | × | √ | × | × | × | × | × | × | × | × | × | × | × | × | × |
| [S25] | GAMBAS | P | × | × | √ | × | √ | × | √ | √ | √ | × | × | × | × | × | × | × | × | × | × |

| Ref | Name | Type | | | | | | | | | | | | | | | | | | | |
|---|---|---|---|---|---|---|---|---|---|---|---|---|---|---|---|---|---|---|---|---|---|---|
| [S26] | SmartCityWare | P | × | × | × | √ | × | × | × | × | × | × | × | × | × | × | √ | × | × | × | × |
| [S27] | Noise mapping | P | × | × | × | × | × | × | × | × | × | × | × | × | × | × | √ | × | × | × | √ |
| [S28] | Vlacheas | P | × | × | × | × | × | × | × | × | √ | × | × | × | √ | × | × | × | × | × | × |
| [S29] | MLSC | P | × | × | × | × | √ | × | × | × | √ | × | × | × | × | × | √ | × | × | × | × |
| [S30] | Yang | P | × | × | × | × | × | × | × | × | × | × | × | × | × | × | × | × | × | × | × |
| [S31] | TMN | C | × | × | × | × | × | × | × | × | × | × | × | × | × | × | × | × | × | × | × |
| [S32] | RERUM | P | × | √ | × | × | √ | × | × | √ | × | × | × | √ | × | √ | √ | × | × | √ | √ |
| [S33] | RAMI | C | × | × | √ | × | × | × | × | × | × | × | × | × | × | × | × | × | × | × | √ |
| [S34] | EPIC | P | × | √ | × | × | √ | × | × | × | × | × | × | × | √ | × | × | × | × | × | √ |
| [S35] | ClouT | P | × | × | × | × | × | × | × | × | × | × | × | × | × | × | × | × | × | × | × |
| [S36] | TSB | M | × | × | × | × | × | × | × | × | × | × | × | × | × | × | × | √ | × | × | × |
| [S37] | EdSC | P | × | × | × | × | × | × | × | × | × | × | × | × | × | √ | × | × | × | × | √ |
| [S38] | SOFIA | P | × | × | × | × | × | × | √ | × | × | × | × | × | × | × | × | × | × | × | × |
| [S39] | CityPulse | P | × | × | × | √ | √ | √ | × | √ | √ | √ | √ | √ | × | √ | √ | √ | × | × | √ |
| [S40] | SCCM | M | × | × | × | × | × | × | × | × | × | × | × | × | × | × | × | × | × | × | √ |
| [S41] | OGC | P | × | × | × | × | × | × | × | × | × | × | × | × | × | × | √ | × | × | √ | × |
| [S42] | PLAY | P | × | × | × | × | × | × | × | × | √ | √ | × | × | √ | × | × | × | × | √ | × |
| [S43] | Nitti | P | × | × | √ | × | × | × | × | × | × | √ | √ | × | × | × | × | × | × | × | × |
| [S44] | BASIS | P | × | × | × | × | √ | × | √ | × | × | × | × | √ | √ | × | × | × | √ | × | √ |
| [S45] | BSI | S | √ | √ | × | × | × | × | × | × | × | × | × | × | × | × | √ | × | √ | × | √ |
| [S46] | SORASC | P | × | × | × | × | × | × | × | × | × | × | × | × | × | × | × | × | × | × | × |
| [S47] | IBM | P | √ | × | × | × | × | × | × | × | × | × | × | × | × | × | × | × | × | √ | × |
| [S48] | ESPRESSO | M | × | × | × | × | × | × | × | × | × | × | × | × | × | × | × | × | √ | × | × |
| [S49] | InterSCity | P | × | × | √ | × | √ | × | × | × | × | × | × | × | × | × | × | × | × | × | × |
| [S50] | ICore | P | × | × | × | × | × | × | × | × | × | × | × | √ | √ | × | × | × | × | × | × |
| [S51] | Agri-IoT | P | × | × | √ | × | √ | × | × | √ | √ | × | × | × | × | × | × | × | × | × | × |
| [S52] | U-City | P | × | × | × | × | × | × | × | × | × | × | × | × | × | × | × | × | × | × | × |
| [S53] | DIAT | P | × | × | × | × | × | × | × | × | × | × | × | √ | √ | × | × | × | × | × | × |
| [S54] | SmartSantander | P | × | × | × | × | × | × | × | × | × | × | × | √ | × | × | × | × | × | × | × |
| [S55] | Collins | M | √ | √ | × | √ | × | × | × | × | × | × | √ | × | √ | × | √ | √ | × | × | × |
| [S56] | Ignite | M | × | √ | × | × | × | × | × | × | × | × | × | × | × | √ | × | × | × | √ | √ |
| [S57] | ACOSO-Meth | M | × | √ | × | × | √ | √ | √ | √ | √ | × | × | √ | × | √ | √ | √ | × | √ | √ |
| [S58] | INTER-METH | M | × | √ | × | × | × | √ | × | × | × | × | × | √ | × | √ | × | √ | √ | √ | √ |
| [S59] | ThingSpeak | P | × | × | √ | √ | √ | √ | √ | √ | √ | × | √ | √ | × | √ | √ | × | × | × | × |
| [S60] | BET | M | × | × | × | × | × | × | × | × | × | × | × | × | × | × | √ | √ | × | × | √ |
| [S61] | Galliot | M | × | × | √ | × | × | × | × | × | × | × | × | × | × | √ | × | × | × | × | × |
| [S62] | Thinger.io | P | × | × | √ | √ | √ | √ | √ | √ | √ | × | √ | √ | × | √ | √ | √ | × | × | × |
| [S63] | IoTEP | P | × | × | × | × | √ | √ | √ | √ | √ | × | √ | √ | × | √ | √ | √ | × | √ | √ |

## 4.3 RQ1: what is the application and type of these approaches?

### 4.3.1 Context

The feature of *context* in the evaluation framework characterises the geographical location and an application area for which an approach is offered.

#### 4.3.1.1 Geographical application

The feature of *geographical application* is based on the fact that an IoT platform may change the ways city services will operate and coordinate. It is important to examine if an approach incorporates the factors related to geographical area in its development process. These factors are, for example, stakeholder's negative attitudes, political background, population diversity, regulations and laws, and city infrastructure readiness. They may slow down or become impediments in the further phases of the development process as mentioned in WSO2 [S23]. We found that the geographical application of approaches, as stated in their documents, is limited to a level of international, continental, national, state, city, suburb, or region as evidenced in Table 4.

Table 4 Geographical applicability of the selected approaches

| Platform | Geographical applicability |
|---|---|
| IoTEP [S63] | International |
| VITAL [S2], CiDAP [S3], IoT-ARM [S8], OpenIoT [S9], FIWARE/OASC [S12], Vilajosana's platform [S13], GAMBAS [S25], RERUM [S32], EPIC [S34], SOFIA [S38], CityPulse [S39], OGC [S41], ESPRESSO [S48], ICore [S50], DIAT [S53], BET [S60] | Europe |
| Padova [S24] | Italy |
| Noise mapping [S27] | Australia |
| Yang's platform [S30] | China |
| RAMI [S33] | German |
| ClouT [S35] | Europe-Japan |
| TSB [S36] | UK |
| Nitti's platform [S43] | Italy |
| BSI [S45] | UK |
| U-City [S52] | Korea |
| Galliot [S61] | Egypt and North Africa |
| RASCP [S1], Khan's platform [S4], Gubbi's platform [S5], Cisco [S6], MC-IoT [S7], Guth's platform [S10], Ganchev's platform [S11], Scallop4SC [S14], Khan' platform [S15], Catherine's platform [S16], EADIC [S17], Telco USN-Platform [S18], SCRM [S19], SPITFIRE [S20], Giang's platform [S21], Wenge's platform [S22], WSO2 [S23], SmartCityWare [S26], Vlacheas's platform [S28], MLSC [S29], TMN [S31], EdSC [S37], SCCM [S40], PLAY [S42], BASIS [S44], SORASC [S46], IBM [S47], InterSCity [S49], Agri-IoT [S51], SmartSantander [S54], Collins [S54], Ignite [S56], ACOSO-Meth [S57], INTER-METH [S58], ThingSpeak [S59], Thinger.io [S62] | Not stated |

### 4.3.2.2 Application domain

As the name implies this feature is to characterise the domain for which an approach and its resultant IoT platform is suitable to use. For example, if the purpose of a platform is to support mission critical services to citizens, the development process should explicitly incorporate additional supportive real-time response mechanisms into the platform architecture. Table 5 shows the classification of the approaches based on their application domains. As shown, MC-IoT [S7] is an architecture for mission-critical IoT based systems where a failure in that system may cause economical and environmental issues. In fact, MC-IoT [S7] is a model driven development process relies on automated model transformation and self-adaptation mechanisms. An application of FIWARE [S12] platform is to implement real e-health remote patient monitoring services. RAMI [S33] describes the crucial aspects of coordination and automation of IoT based manufacturing systems regarding Industry 4.0. OGC [S41] is developed to access and integrate different sources of geospatial information for smart cities. Nitti's platform [S43] has been developed for the sustainable tourism domain in order to optimise the movement of cruise ship tourists in cities regarding factors such as transport information and queue waiting times. BASIS [S44] is pertinent to the flight itinerary applications where it provides services to search and visualise delay profiles in flights of cities due to, for example, weather conditions. Agri-IoT [S51] is a platform developed for the smart farming in the food supply chain. Documentation of ThingSpeak [S59] shows the platform is developed for more home/private use applications (e.g. monitoring the temperature and humidity of an office) rather than organizational.

Table 5 Application domain of identified approaches

| Platform | Application domain |
|---|---|
| MC-IoT [S7] | Mission-critical systems |
| FIWARE/OASC [S12] | Retail, healthcare, logistic, agriculture |
| Scallop4SC [S14] | Large-scale log data processing |
| Telco USN-Platform [S18] | Automobile |
| MLSC [S29] | Smart health system |
| [S30] | Polytrophic power |
| RAMI [S33] | Manufacturing |
| CityPulse [S39] | Real-time large-scale data processing |
| OGC [S41] | Geospatial |
| PLAY [S42] | Social media |
| [S43] | Tourism |
| BASIS [S44] | Airline transport |
| Agri-IoT [S51] | Agriculture |
| ThingSpeak [S59], BET [S60], IoTEP [S63] | Smart energy |
| RASCP [S1], VITAL [S2], CiDAP [S3], Khan [S4], Gubbi's platform [S5], Cisco [S6], IoT-ARM [S8], OpenIoT [S9], Guth [S10], Ganchev [S11], Vilajosana [S13], Khan [S15], Mulligan [S16], EADIC [S17], SCRM [S19], SPITFIRE [S20], [S21], [S22], WSO2 [S23], Padova [S24], GAMBAS [S25], SmartCityWare [S26], Noise mapping [S27], [S28], TMN [S31], RERUM [S32], EPIC [S34], ClouT [S35], TSB [S36], EdSC [S37], SOFIA [S38], SCCM [S40], BSI [S45], SORASC [S46], IBM [S47], ESPRESSO [S48], InterSCity [S49], iCore [S50], U-City [S52], DIAT [S53], SmartSantander [S54], Galliot [S61], Thinger.io [S62] | Not stated |

## 4.4 RQ2: How IoT platform development process lifecycle is perceived in the literature?

Derived from the identified approaches, the relations between IoT development phases is shown in Figure 4. It is important to realise that the common software system quality factors such as interoperability, security, reusability, configurability, energy efficiency should be viewed as crosscutting concerns across different layers and development process of an IoT platform.

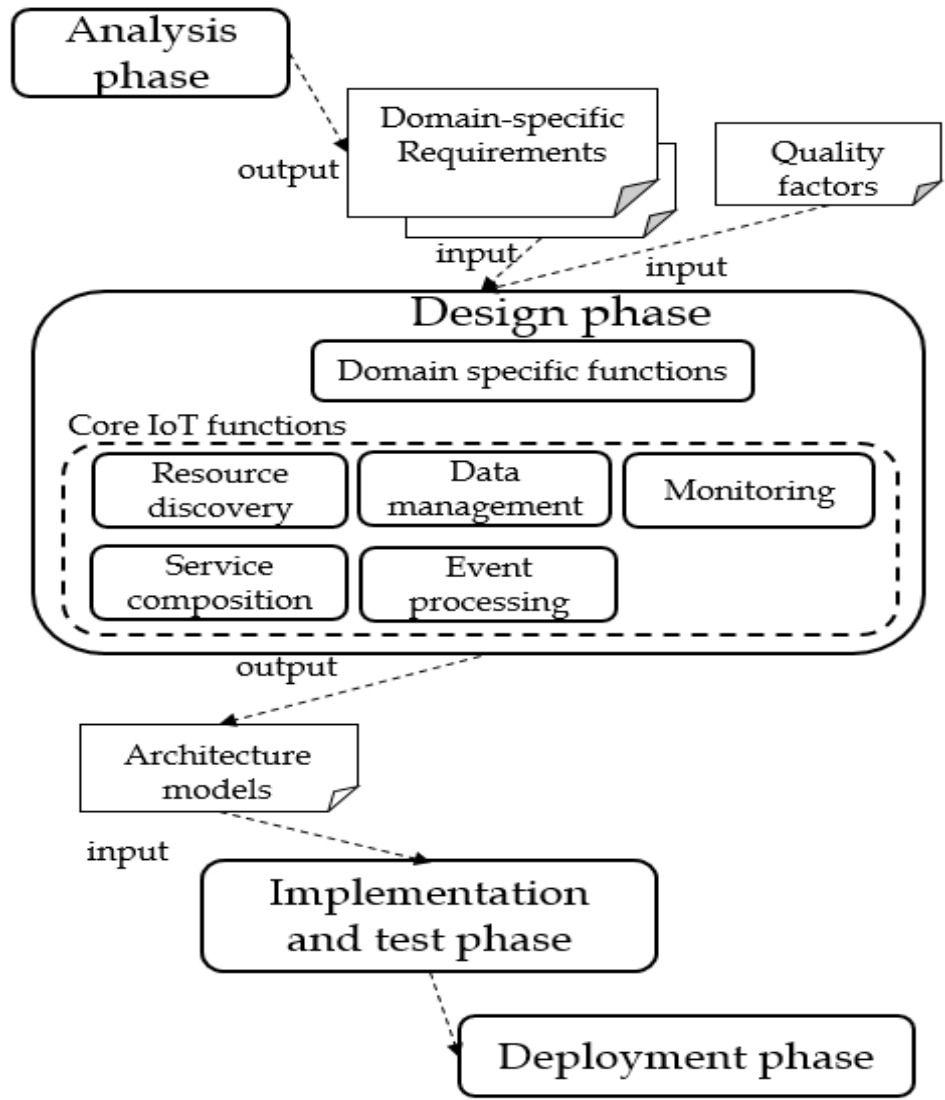

Figure 4. An overall process of IoT platform development derived from the approaches. In the design phase, identified requirements are used to design logical high-level IoT platform architecture models realizing core IoT and domain specific functions. The architecture models are operationalised in the implementation and test phase and finally deployed.

### 4.4.1 Initialization phase

The purpose of this phase is to establish a project plan and to analyse the feasibility of the IoT technology to operationalise a citizen-centric vision. According to IoT-ARM [S8] and BSI [S45], this phase defines a clear and compelling city vision (what good looks like) for the platform development in the subsequent phases. Amongst other things, a city vision document may explain objectives to achieve critical services that are to offer by the platform to citizens. Defining the city vision is an iterative and collaborative task and it needs an active participation of all smart city stakeholders. BSI [S45] defines the following recommendations when planning an IoT project:

— establishing common terminologies, i.e. project glossary, to ensure all stakeholders have a clear, consistent and common understanding of the key concepts involved in the platform development;
— acquiring right skills and interdisciplinary team arrangement;
— managing probable organizational/city change;
— deploying a transparent governance process to monitor platform development.

Apart from that, a key concern that should be planned ahead and further elaborated in the later phases is to address the interoperability across the IoT platform components.

Two approaches SCRM [S19] and BSI [S45] emphasise defining an individual plan defining integration strategies and innovative characteristics contributing to a green and sustainable city. This helps identifying key barriers and corresponding solutions to achieve the interoperability quality factor.

### 4.4.2 Analysis phase

This phase aims at the specifying and prioritizing functional requirements as well as the quality factors that should be realized by the target IoT platform. The stakeholders' requirements depend on the target context (i.e. the features of *geographical application* and *application domain*) chosen to build the platform.

We found that the existing approaches prescribe using the common requirements engineering techniques available in traditional software engineering literature to elicit requirements, for example, using object-oriented use-case modelling as suggested in IoT ARM [S8] and ACOSO-Meth [S57]. A user interface centric requirement analysis technique is proposed by Ignite [S56] through which a prototype of IoT functions exposing to end users including actions and views are generated. An action can be a trigger like a button or slider whilst a view is a layout of the system such as tables, diagrams, or textual layers. Together views and actions indicate business logic and user permissions to work with IoT services. In Collins [S55] there is a stream of requirements analysis entitled as the infrastructure analysis. The reason for doing this analysis is to know how new IoT applications and services will be deployed and integrated with the existing IoT infrastructure components such as hardware, database, and middleware.

As far as the quality factors are concerned, much effort in the analysis phase is spent on the analysis of security and interoperability. Developers should elicit and identify correct security requirements for each layer of the platform from the low-level networking perspective to end-user level. Identifying security requirements at the early stages of the IoT platform development is crucial as smart cities consist of a wide range of heterogeneous smart objects and technologies that dynamically join and leave the network. Such ever-changing environments may raise unforeseen security and privacy risks. EPIC [S34] suggests the security requirements can be identified through the following steps:

— analysing behaviour and communication among smart objects, platform components, and humans involved in the smart city context and;
— determining what/when/how different types of data should be protected.

Given the heterogeneity of platform components or platforms with together which may be integrated with respect to certain goals, interoperability requirements should be identified for each layer of the platform. In this regard, INTER-METH [S58] recommends taking the following steps for interoperability requirements analysis:

— identifying integration points across the platform layers;
— identifying requirements of each integration point;
— writing objective statements for each integration point to help developers focus their efforts during design and implementation phase.

### 4.4.3 Design phase

The feature of *design* in the evaluation framework examines if an approach addresses the core functions of an IoT platforms. Based on the reviewed approaches, these core functions, which defined as sub-features in the evaluation framework, are *resource discovery*, *data management*, *monitoring*, *service composition*, and *event processing*. They provide a backbone for an IoT platform to address domain specific user requirements and quality factors. The following subsections present the results of the evaluation of the existing approaches against these five features.

#### 4.4.3.1. Resource discovery

In a dynamic environment of a smart city, smart objects can continuously join and leave the network. A key expected function in an IoT architecture is that smart objects should be discoverable or connectable to the platform in both automatic and manual ways. In general, this needs implementing the following mechanisms in the platform:

—defining unique identifiers for smart objects;
—enabling smart objects to announce their presence in the network and register themselves;
—enabling users to discover, browse, and perform queries over objects in the network span.

In approaches such as OpenIoT [S9], Telco USN-Platform [S18], GAMBAS [S25], VITAL [S2], and FIWARE [S12], it can be seen that *publish/subscribe* is a common mechanism used for the resource discovery function as shown in Figure 5. In this mechanism, a middleware component, also called distributed registry, provides necessary APIs and enables smart objects to register themselves in the platform network. The registration makes them discoverable by other smart objects and software components and enable them to upload/post and disseminate data/meta-data to the rest of the network. In addition, the middleware's APIs should allow users running queries over the inserted data in the network by smart objects, seeking other subscribed smart objects in the network, browsing the list of available services and resources proposed by smart objects deployed by other users regarding their locations. In order to explore services offered by smart objects subscribed in the network, the semantic service matchmaking mechanism is used. An implementation of this mechanism suggested by IoT-ARM [S8] where a lightweight description ontology enables users or software components to search IoT services. Similarly, MC-IoT [S7] suggests a model-driven transformation mechanism for identifying, specifying, realizing, and composing new resources and services.

The interoperability and security are two important quality factors that should be taken into account. As far as the interoperability concerned, smart objects may not be detectable or searchable due to incompatibility with other smart objects. In other words, a smart object may want to send a signal showing its availability to another smart object but both have different interfaces. In addressing this issue, Nitti [S43] suggests using *virtual objects* (VO) mechanism. In this mechanism, VOs are digital counterpart models of physical objects which encapsulate information and operations of physical objects. A VO applies the separation of concern principle to hide incompatibilities where it makes logical links between the real-world objects and relevant virtual objects. The platform has a search and discovery engine component that receives service requests from users. Requests are compared with virtual templates of VOs to discover the most similar and available VO instances. In similar way, VITAL [S2] allows users defining an abstraction layer through which a VO handler points to physical items which can be discovered, selected, or removed. Alternatively, Thinger.io [S62], defines client libraries so that smart objects can connect to the Thinger.io platform, use efficient bidirectional communications, and consume from any external application.

With respect to the security, Nitti [S43] suggests defining three levels of VO discoverability as follows:
— *public* where a registered object can be discoverable to all users in the network;
— *private* where a registered object is discoverable by the object owner;
— *friend* where a registered object is discoverable by a private key provided by the virtual object.

An issues related to the resource discovery function is the possibility of having duplicated identifiers for different smart objects that connect to the network. It is likely that some objects from different platforms have same identifiers with those objects that have already been connected to, and setup by, different platforms. To have smart objects with a unique identifier, a platform should define a naming mechanism for those joining the network. In CiDAP [S3], a uniform resource name (URN) is implemented where some predefined prefix such as location, name, and identifier are added to objects. Prefixes should be kept short to reduce processing overhead.

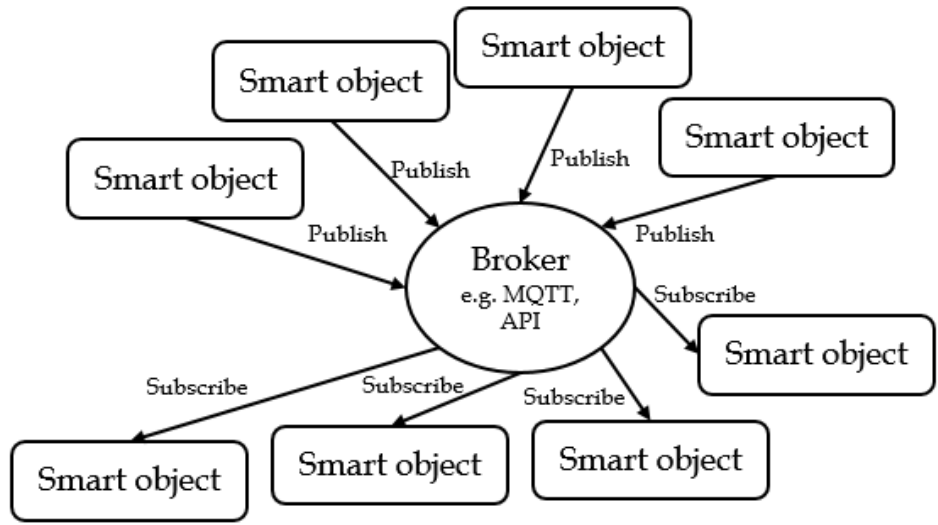

Figure 5. *publish/subscribe* mechanism proposed by the existing approaches for resource discovery—adapted from OpenIoT [S9], Telco USN-Platform [S18], GAMBAS [S25], VITAL [S2], and FIWARE [S12]

### 4.4.3.2. Data management

**Data collection.** A primary function of an IoT platform is to collect data from various sources in the environment as well as across all its layers. The data, which can be classified as semi-structured and structured as stated by CiDAP [S3], two approaches RASCP [S1] and Scallop4SC [S14] define three sources of data that should be continuously captured and stored by a platform:

— data about physical city entities such as smart objects, citizens, traffic model, sensor network model, data model, city maps, and energy distribution model;
— data about working software components, services, and applications including source codes/libraries and associated documents;
— historical data such as log data, censor states, citizens' action history.

In terms of the data type, Thinger.io [S62] defines three classes of data for collection such as observational data (i.e. original data about dynamic scenarios as collected from heterogeneous objects), contextual data (i.e. data about circumstances of objects), and knowledge models (i.e. a priori or inductively learned).

The designing a data collection function involves with addressing interoperability, security, scalability, and configurability as described in the following. An IoT platform needs to collect data from heterogeneous smart objects and applications each using different formats to store and share data. Hence, as recommended in Khan [S4], an appropriate APIs that either are provided by data source providers or the platform provider play an important role for the quality of the data acquisition from the data sources. This assists users of the platform to perform sophisticated queries to extract data for data processing purpose. A commonly software engineering mechanism for the data collection function is using adaptors/wrappers as shown in VITAL [S2], CiDAP [S3], and OpenIT [S9]. More exactly, OpenIT [S9] has a component called Extended Global Sensor Networks (E-GSN) which collects data via serial port communication of sensors, HTTP requests, and JDBC (Java DataBase Connectivity) queries. The E-GSN implements wrappers for both data providers and developers to implement their own customised data acquisition functions. Similarly, the approach CiDAP [S3] suggests defining a unified interface for the data collection including two main components named IoT-broker and IoT-agent. An IoT-agent connects to sensors and responds to requests from the IoT-broker on behalf of real sensors. The IoT-broker forwards requests to IoT-agents and pushes returned results back to data storages. VITAL [S2] has components named Virtualized Unified Access Interfaces (VUAIs). A VUAI implements a collection of connectors and drivers to enable communication with other platforms. The connectors use linked data standards such as RDF, JSON-LD, and ontologies to represent data format and data access.

The data collection from the data sources should be performed in a secure way in the sense that the data traversing in the network should be encrypted in the sender and decrypted on the receiver side via cryptography algorithms. The following basic mechanisms are suggested by the GAMBAS [S25]:

— authentication and authorization for data access;
— data encryption via cryptographic algorithms through which the traversing data are encrypted and decrypted in the receiver side;
— secure protocols such as TLS/SSL or IPSec at network layer;
— establishing regulations and rules in favour of citizens' privacy and security;
— hardware obfuscation where functional logic of sensors coded to prevent reverse engineering attacks;
— minimising data acquisition from smart objects.

Regarding the configurability quality factor, the data collection function should be designed to address two changeable modes namely *real-time* or *near-time* indicating the extent of being real-time for the data collection. Such a view, is defined in OpenIoT [S9] which provides a publish/subscribe middleware to allow users to set a data acquisition mode.

In addition, the data accumulation function can be concerned in terms of network traffic and performance quality factors. That is, malicious users may connect to the network using smart objects (e.g. mobile devices) and send continuous data or service request inquires. OpenIoT [S9] implements a mobile broker running on smart objects which prevents potential data overload and ensures only relevant data is transferred from objects to the platform.

As the data collection functions performing on the smart objects are typically battery powered, an important quality factor that should be addressed in the infrastructure layer of a platform is the energy efficiency. Hence, the data acquisition should be resource efficient as suggested by GAMBAS [S25]. For example, Thinger.io [S62] uses a mechanism called Protoson for transferring data between smart objects and platform's servers/software components. Unlike HTTP approach for the data transfer, which includes several JSON or XML headers and payloads, Protoson uses raw compact binary connections without the HTTP overhead. Thinger.io [S62] shows that Protoson mechanism saves bandwidth and reduces power consumption in smart objects.

**Data cleaning.** This function is to remove anomalies from data prior to storing them into databases. The data cleaning function can be considered from the perspective of reliability quality factor, i.e. sufficiently completed and error-free data. The data collected from sensors might be noisy and abnormal which may affect the reliability of derived data analysis results. For example, a sensor may report a temperature which is out of the expected range or may stop report. This might be due to several reasons such as a battery run out or unexpected broken network repeater. Anomaly detection algorithms should be designed as recommended by CiDAP [S3]. In the view of Cisco platform [S6], the data cleaning function should include the following generic steps such as:

— reconciling data formats collected from data sources;
— ensuring the semantic consistency of data;
— normalising and de-normalising the data to get faster process.

In terms of the configurability quality factor, not all generated data in smart city environment may be the equal interest of platform's users. The data cleaning function should allow users to store and process a sub set of data that meet their requirements, for example, a specific threshold value or comparison measured value. Hence, the platforms such as VITAL [S2], OpenIT [S9], CityPulse [S39], and SOFIA [S48] define a context-filtering component which continuously monitor environments to automatically select events subjected to users' interests and to create user-specific data filtering patterns. For example, VITAL's [S2] filtering component enables to collect the data which meets requested a data interpolation pattern. Moreover, in OpenIT [S9] the component CloUd-based publish/subscribe middleware facilitates the filtering of data streams e.g. sensor data in a way that only data that are subjected to the interest of users are collected which also avoids potential data overload.

**Data storing.** This function is to manage storing collected data in the physical data storages of the platform. Due to the variety and large volume of data, the majority of the existing approaches suggest two types of data storages:

— relational databases that are a common option if atomicity, consistency, isolation, durability (ACID) constraints and support for complicated queries required;

— No-SQL databases such as Hadoop, CouchDB, CouchBase, MongoDB, and HBase databases supporting features such as horizontal scalability, distributed index, and dynamically modifying data schema.

**Data processing.** This function provides sophisticated data analysis over the collected data. The function leverages APIs and data mining techniques for classification, regression, and clustering that are supported by data analytics platforms such Apache Storm, Apache Spark, and Hadoop MapReduce. The existing approaches include two modes of data processing namely real-time stream processing and batch processing. The data processing function is concerned for the scalability quality factor to respond to data processing requests effectively. Need for addressing the scalability comes into play when the is an increase in number of sensors, data volume, communications among objects, processing demands, and users connected to the platforms as discussed in PLAY [S42], Nitti [S43], CityPulse [S39], and DIAT [S53]. The scalability in the existing approaches leverages enabling technologies such as cloud computing, data analytics, and micro-services. For example, Khan [S4] uses Hadoop MapReduce to scale the data processing function in its platform. ThingSpeak [S59] provides APIs to write and execute code to process data via the proprietary Matlab tool, which may impede the popularization of the platform.

**Query processing.** A query processor function performs queries over the platform data storages. Similar to the resource discovery function, the common mechanism used in the approaches for the query processing function is the publish/subscribe. An example of that is CityModel API suggested by CiDAP [S3] which is hosted on servers and it allows users to subscribe and perform queries. A simple query is to request a real-time snapshot of data over all databases or smart objects whilst a complex query is to request aggregated results over the historical data collected within a period. To keep users notified of the latest data all the time, the CityModel API also defines two types of the subscription mechanisms called CacheDataSub (in the database) and DeviceDataSub (in objects) in order to provide different expected latency.

**Meta-data generation.** The purpose of this function is to improve the classification, identification, decision making, and retrieval of the data from the data storages and smart objects. For example, in CityPlus [S39] the characteristics of sensors such as its location, the interval of updates for data fetch, and data category are described using sensory meta-data which is named SensorDescription. Meta-data can be either generated manually by users through the provided platform's user interfaces or by the platform internally during a data collection and processing. Meta-data generation function is in relation to the interoperability quality factor. That is, via using the meta-data, smart objects can identify, communicate, and exchange data with together by seeking their models, types, and other attributes. In other words, the meta-data acts like a guideline that helps smart objects to process the counterpart data correctly.

**Data visualisation.** In a platform, this function is responsible to provide interactivity and user interfaces to enable users to send queries on topic of interests and to get graphical view of the data analysis results. This can be in the forms of mobile applications, dashboards, reports, message boards, 3D spaces, and 2D maps. For instance, ThingSpeak's API's [S59] enables users to visualize collected data through using spline charts.

### 4.4.3.3. Monitoring

An IoT platform should ensure the satisfaction of the quality factors for both internal environment, i.e. its components, and external environment, i.e. smart city. The monitoring function is realized in an architecture by components and APIs for keeping the track of the environment and subsequently performing resource allocation and task scheduling algorithms. Scallop4SC [S14], SmartCityWare [S26], and RERUM [S32] define the different types of log data that should be analysed to detect anomalies in the environment as follows:

— *system log* to capture the history of operations and resource usage of components;
— *energy log* to capture the history of energy consumption by smart objects deployed in different regions and their battery lifetime;
— *device log* to capture the history of their operations, status e.g. changing TV channel, and switching on/off light;
— *network log* to capture link quality, throughput and delay, transmission queue size, the number of collisions, packet error rate and other critical networking statistics;
— *environment log* to capture the history of changes in the environment such as temperature, humidity, and the number of people

The *CityPulse* [S39] suggests defining a component responsible for comparing the collected data streams, identifying anomalies, correlations, or similar patterns of divergence and generating alerts in the case of occurrence abnormal behaviours in the environment.

To implement the monitoring function, FIWARE [S12] suggests two approaches: distributed and central. In central approach, a set of monitoring components are installed on server and gateway collects statistical data from all smart objects in the network, analyses, and identifies malfunctions and failures. A disadvantage of this approach is high-energy consumption and heavy signalling in the network and data transfer. On the contrary, in the distributed monitoring approach, each smart object monitors itself and its neighbour smart objects to find issues in the network. The distributed approach although does not create high signalling in the network, it causes more energy consumption on smart objects because each object should perform complex monitoring algorithms.

From the reviewed approaches, we found two other quality factors related to the monitoring function: interoperability and scalability. Firstly, the monitoring may not be applicable in heterogeneous smart city environment as smart objects may use different physical layers and networking technologies. To address this issue, FIWARE [S12] suggests providing a set of APIs for developers which wrap incompatibilities between platform and smart objects. This enables developers to manage aggregated and real-time monitoring data. Secondly, an IoT platform should define mechanisms to keep expected performance without negatively affecting the quality of existing services in peak time when the number of smart objects connecting to the platform is increased. This is related to the scalability quality factor and it is supported in Noise mapping [S27] and IBM [S47].

### 4.4.3.4. Service composition

All individual services in an IoT platform that are offered by smart objects or platform components can be combined to create large services. These new composite services can be deployed and executed on top of the IoT platform. If an IoT platform allow developers for an end-to-end cross-platform development of composite services, the cost for building new IoT based applications is reduced. The logic for defining a specific service composition layer in some existing approaches such as SCRM [S19], TMN [S31], RERUM [S32], ClouT [Ss35], and OGC [S41] is to enable users in creation of new IoT applications via combining existing services. The service composition function in Meta Services of VITAL [S2], generic enabler in FIWARE [S12], Software-as-a-service (SaaS) in WSO2 [S23], and IoT-ARM [S8] commonly rely on the service-oriented architecture (SOA) approach. For example, WSO2 [S23] provides the WSO2 Private PaaS (platform as a service) product which is based on Apache Stratos project to enable developers in building scalable applications. FIWARE [S12] provides a set of pre-built general-purpose functions accessible through APIs, called Generic Enablers (GEs), which allow developers to build new applications to run on platform. When viewed collectively, a service-oriented approach to define the service composition function in a platform should have the following sub-functions:

— A service orchestration function enabling users to define inputs/outputs and business rules (e.g. sensor data entry validation, process sequences, or authorization) in order to create new services by selecting and combining pre-existing IoT services;

— A service choreography function offering a broker to handle publish/subscribe communication between services.

To enable service composition function in an IoT platform, the quality factors such as interoperability, reusability, security, and energy efficiency are concerned as elaborated in the following.

A mechanism to address the interoperability, as already mentioned for the resource discovery function in Section 4.4.3.1 is VO suggested. This mechanism is implemented in Vlacheas [S28] and iCore [S50]. VO introduces the composite virtual objects (CVO), a cognitive mashup of semantically interoperable virtual objects and their offered services. We also identified a set of design principles for addressing the interoperability. For instance, SmartCityWare [S26], EdSC [S37], and Nitti [S43] suggest that platform components should be loosely coupled so that they can simply be integrated with other components. In addition, ESPRESSO [S48] suggests the following recommendations to support the service interoperability:

— implementing small and reusable services (e.g. microservices) and APIs with minimum functions instead of large and coarse-grained services;
— using common standards for data exchange such as IP-v4, IP-v6, IP-Sec;
— offering data processing services instead of data exchange services (avoiding offering data as a service).

In addition, at the infrastructure layer of a platform including hardware, data centres, networks, and devices, the approaches commonly use de-facto standards and protocols, such as HTTP, MQTT (Message Queuing Telemetry Transport which is a broker-based publishing/subscribing), and AMQP (Advanced Message Queuing Protocol) to address interoperability. They are middleware protocols which extensively used for exchanging messages across among different objects at the network layer.

Another quality factor in relation to the service composition is that a platform should provide sufficient support for service reusability. In Nitti [S43], the reusability is used for the purpose of the data processing function where the data are collected from various sources but they are processed and used in a similar way. Therefore, the data processing of the platform can be reused with minor modifications for constructing new processing instances. The common approach to improve service reusability in the existing approaches such as IoT-A [S8], SPITFIRE [S20], iCore [31], RERUM [S32], ClouT [S35], TSB [S36], and DIAT [S53] is using of the model-driven development in which a base architecture, i.e. reference architecture, with minimum core services, and technologies is designed which is extended and reused for new service composition. The security should be taken into account when creating and using composite services in a way that a composite service should still satisfy the expected security requirements with an acceptable accuracy.

Finally, a composed service may be based on invoking the fine-grained services of smart objects that have low or little storage/computational power, middle-end with restricted resources, i.e. sensors, to high-end, i.e., smart phones and laptops. Finally, as far as energy efficiency concerned, a common advice by the existing approaches is to consider resource constraints, e.g. battery, of involved objects when making composite services.

### 4.4.3.5. Event processing

This function in a platform aimed at representation, capturing, and quickly react to important events either external environment such as city events, peak-time vehicle speed, geographic events as well as internal environment for example component events. An event is an observable change in the state of the environment which can be triggered by a stimulus e.g. sensors. Event processing should be conducted in a real-time way so that users can receive accurate and timely response. According to EdSC [S37], the basic functions to support being event-driven in a platform are:

— *signal-to-event* to convert signals in the environment to meaningful events which can be used by smart objects or human;

- *knowledge sharing* to represent events;
- *action definition* to store event condition and perform actions in the case of occurrence of event incidence;
- *action-to-signals* to convert functions performing in smart objects into signals.

To address customisability, Galliot [S61] suggests that a platform should enable users to write and execute their specific codes once an event occurs. Cisco [S6] adds some other functions such as sampling/filtering events, comparing events, and joining events to create complex events.

Exchanging and processing heterogeneous events coming from multiple sources at different levels of the platform stack is a challenging issue which should be addressed during an architecture design. Our review revealed that the technique to address event interoperability in the approaches, such as PLAY [10] and CityPlus [S39], is based on using ontologies, which provide a common semantic basis for data and metadata representation and interpretation. That is, events that are represented using ontology concepts and annotated via metadata enable their correct interpretation by other smart objects or services in the network.

### 4.4.4 Implementation and test phase

This phase focuses on the implementation of the designed architecture in previous phase. In the existing approaches such as Collins [S55], Ignite [S56], ACOSO-Meth [S57], and INTER-METH [S58], this phase is to implement and test three classes of components, each with specific functionalities as described in the following:

**Software layer components** to enable users to perform platform's functions as explained in the design phase (Section 4.4.3). These components are, for example, software applications, services, ERP (enterprise resource planning), mobile applications, business analytics applications/reports, back-end services, and monitoring applications. The components in this layer are responsible for:

- receiving data from smart objects, performing processing algorithms over the data and sending the results back to end users and smart objects;
- providing an end-to-end application development foundation to build new applications to be run on top of the platform;
- providing APIs to extend the platform with new services;
- orchestrating and managing business processes, services, and applications.

To implement software components, the approaches use the programming languages such as C, C++, Java, and Python and machine-level languages such as C++ and Assembly for low power devices. For instance, ThingSpeak's APIs [S59] provides a development environment to write data processing functions via programming languages such Ruby, Python and Node.js.

**Data layer components** to store and retrieve data from/to physical data storages.

**Infrastructure layer components** to provide hardware and resources for data storages, computations, and physical interconnectivity among data centres, servers, and networks.

**Smart objects layer components** to collect data from the physical environment and send/receive the data to/from software components. Coding the smart objects may require dozens of line of codes such as managing HTTP requests, adding headers, sending gathered data, parsing input/output, commanding, and allocating resources. Thinger.io [S62] provides client libraries to simplify coding smart objects. We found that the following enabling technologies are used in the existing approaches:

- cloud computing to provide a scalable and highly unlimited resources to collect and store data from devices and to perform data intensive analysis;
- fog computing to provide low and predictable latency and geographical IoT distributed applications;
- data analytics platforms to manipulate collected data set;
- restful services to enable coordination and composition of IoT services;
- VO to integrate IoT sources such as smart objects.

A key quality factor regarding this phase is the reusability of platform components. There is a possibility to develop new platforms or to extend existing ones using pre-existing IoT platforms. In this regard, VITAL [S2] provides a set of visual tools and capabilities which allow a rapid development and deployment of new IoT applications and back-end services based on NodeRED (nodered.org) open source development tool proposed by IBM platform [S47]. OpenIoT [S29] has a visual integrated development environment that accelerates building and deploying new IoT applications. The platform reduces programming effort by providing options such as visual sensor discovery regarding their locations, types, configuring and monitoring sensors, and composing services based on Web 2.0 mashups. FIWARE [S12] platform has been built based on SaaS delivery model. Such a foundation including a collection of APIs minimises the developing IoT applications. IoTEP [S63] has been developed based on FIWARE [S12].

Testing is a key development activity to ensure that the implemented platform satisfies the expected functionalities in a target environment. For example, via testing, developers can estimate workload and identify an upper and lower workload that can be processed by different platform deployment configurations. The testing of IoT platforms can be performed at three levels: beginning from the lowest level where each platform component is tested, i.e. unit testing, verifying if platform components work together well, i.e. integration testing, and finally testing the whole platform. Traditional software engineering testing techniques can be employed as it can be seen in Collins [S55] and INTER-METH [S58], but there are some testing challenges. For example, the dynamicity of the environment where different smart objects join and leave the network makes it difficult to conduct all testing scenarios completely. Moreover, an IoT platform may constitute many different components at software, data, and infrastructure layers which becomes difficult to ensure that the whole platform is working effectively.

### 4.4.5 Deployment phase

The deployment phase is complex and it comprises both technical and non-technical concerns. From the non-technical point of view, in particular financial, Vilajosana [S13] recommends a three-stage deployment strategy as follows:
— deploying components that not only offer utility but also offer very clear return on investment and generate cash flows for new investments;
—deploying components that may not necessarily produce direct financial benefit but longer return on investment;
— deploying components offering by third party developers to be added on the top of existing platform to make it self-sustainable through standardized APIs.

Technically speaking, the phase is performed to deploy all components related to the software, data and infrastructure layers to make continuous and close feedback loop. For the software and data components, a concern in this phase is to identify an optimum distribution of software platform components on servers, typically cloud and fog servers, by taking into account the quality factors such as performance and security. Three types of component deployment are suggested by BET approach [S60]:
— all in cloud: deploy all components including processing and storage on cloud servers;
— all in fog: deploy all components including processing and storage on local fog nodes to reduce latency and traffic network;
— half in fog: deploy some components on fog to reduce latency and bandwidth utilization but also benefits from cloud servers for computational power.
In addition, Inter-Meth [S58] defines the following configuration tasks in this phase:
— software component configuration including graphic tools for service orchestration and underlying interoperability mechanism;
— semantics configuration to manages all the processes and mechanisms;
— user configuration to manage authorized access to the IoT platform's resources.
The deployment of infrastructure layer components such as data centres, servers, networks, and smart devices are also concerned in this phase. The existing approaches highlight different tasks to be performed . These include:
— identifying areas where Wi-Fi deployment will result in the sufficient improvement of the service delivery cost TSE [S36],

— an optimum distribution of smart objects to get less traffic congestion and lower round trip ACOSO-Meth [S57],
— generating deployment scripts on smart objects Collins [S55],
— deploying communication technologies and APIs on smart objects ESPRESSO [S48],
— gateway and network configuration with a focus on the interoperability of platform components and smart objects Inter-Meth [S58].

Apart from that, we found the approaches commonly incorporate well-known software engineering design principles such as adoption of open standards (Khan [S4]), loose coupling (SmartCityWare [S26]), (Nitti [S43]), data minimization (RERUM [S32]), open APIs, a-synchronous/synchronous communication, stateless services, and decentralised evolution (Nitti [S43]), and modularity (ESPRESSO [S48]).

## 4.5 RQ3: What roles are involved in the development of IoT platforms?

So far, our discussion on the lifecycle aspect (Section 4.4) has implicitly referred to various roles participating in the IoT development process. As an IoT development process relies on a variety of technical expertise, business acuity, and delivery skill set, there is a need to have a plan for acquiring development roles. This section explicates roles and their associated responsibilities in the view of the existing approaches. According to the Gartner's report: "engagement skills are a fundamental requirement for delivering IoT — and are a critical competency for IoT architects" (Pettey, 2018). A managed role engagement and effective collaboration programme, as defined in BSI [S45], is an important part of IoT projects. For example, an IoT-based city roadmap cannot be effectively produced without an active involvement of roles such as IoT architect and citizens. The role engagement ensures that they have a clear understanding of the IoT smart city program. In particular, IoT architect is responsible for making an engagement between development teams and stakeholders to develop clear business objectives for IoT solutions and to ensure they integrate well together. We found that, except for a few approaches namely SCRM [S19], BASIS [S44], ESPRESSO [S48], and INTER-METH [S58], the majority of the existing IoT approaches do not elaborate on the role definitions as a part of their suggested platform development process. We will discuss this issue further in Section 5.3. Table 6 shows the identified roles along with their corresponding responsibilities.

Table 6. Roles involved during IoT platform development

| Role | Responsibilities | Approach |
|---|---|---|
| Project manager | Initiating, planning, executing, monitoring, controlling IoT project | All |
| IoT architect | Identifying and modelling a target IoT architecture which meet stakeholders requirements | Al |
| IoT programmer | Implementing APIs for providing interoperability, coding, configuring smart objects at machine level | All |
| Third party programmer | Implementing and supporting of third party services | BASIS [S44] |
| Data analyst /data scientist | Designing and implementing data models for the data layer of the architecture | BASIS [S44] |
| Non-relational data storage specialist | Implementing and managing non-SQL related technologies | BASIS [S44] |
| Infrastructure administrator | Procuring, managing, and monitoring physical platform infrastructure | BASIS [S44] |
| Security specialist | Implementing mechanisms to ensure the platform's privacy, security, and integrity | BASIS [S44] |
| Integrator | Identifying integration points and implementing integration layers in order to address interoperability issues | INTER-METH [S58] |

| City leaders/planner | understanding of smart cities, define a development trajectory of a smart city vision towards an IoT platform, and monitoring the smart city projects against the critical success factors | SCRM [S19] |
|---|---|---|
| Citizens | Sharing data | ESPRESSO [S48] |

## 4.6 RQ4: What modelling activities and modelling languages are defined during IoT platform development?

The evaluation framework defines *work-products (artefacts)* and *modelling language* to assess if the modelling aspects is supported in an approach. These features are discussed in the following subsections.

### 4.6.1 Models (work-products/artefacts)

The development activities of an IoT platform produce a series of implicit/explicit models to represent concepts and information. The models help trace how requirements of users are realized in the implemented platform. We observed that the approaches specify models that are either originated from the traditional software engineering or pertinent to the IoT context. We identified a group of models defined by the approaches as presented in Table 7 and briefly described in the following.

Some models are the output of the development activities in the initialization and analysis phases of the development process. In this regard, BSI [S45] prescribes to generate a *smart city road map* as one of the initial models   in an IoT platform development process. The road map model describes city transformation and cover important items including stakeholder collaboration work-stream, leadership and governance processes, and strategies for procurement, supplier management, and risk management. In addition, a useful collection of models related to the early stage of platform development is *city models* such as a traffic model, sensor network model, data model, city maps, city visions, and an energy distribution model, 2D or 3D map. All of these models helps to identify smart objects and required components in the platform. Moreover, *requirement models* represent functional requirements and quality factors that are expected to be realized by a target platform. RERUM [S32] suggests generating the use-case diagrams to represent requirements. Colin's platform [S55] defines a work-product named *IoT Canvas* which is produced during brainstorming sessions conducted by developers and stakeholders in order to identify and validate high-level requirements to be addressed by an IoT platform. Figure 6 shows a typical IoT Canvas including sections such as smart objects, users, data models, and middleware.

| THINGS | END POINTS | MIDDLEWARE | AUTOMATION | USERS |
|---|---|---|---|---|
| Water barrel/pump<br><br>Weather station (*Alecto WS-5000*)<br><br>Solar panel(*Solar log*)<br><br>Washing Machine (Beko)<br><br>Dryer<br><br>Dishwasher | Water level sensor<br><br>Valve control<br><br>Allnet logger<br><br>Smart-Relay box | Raspberry PI<br>XBee Gateway<br><br>Messaging Broker | waterBarrel>90% && *solar panel* >90%,<br>*valve=1&&washer =1*<br>*solar panel* >90%<br>*<appliance>Power =1&&program=1* | House owner (Head geek)<br><br>Family members<br><br>Community members from weather websites |
| **DATA MODEL** | **THIRD PARTY SERVICES** | **WIDGET** | | |
| Valve control-integer<br>Weather station-complex<br>Solar-complex<br>Appliance-boolean | Wunderground<br><br>Personal email | Weather, water, solar - **Graphs**<br><br>Appliance control - **toggle status** | | |

**DESCRIPTION**

"I'd like to make an automation project that "senses" the weather outside (rain, sun radiation and darkness), takes into account the electricity produced by the solar panels and that than automatizes certain household appliances or the central heating.

I would like to have such a system because I want our house to be smarter and less energy-consuming and thus more environmentally friendly."

Figure 6. *IoT Canvas* model for identifying high-level requirements— source Colin [S55]

Some models are produced during the design phase. For example, an *IoT domain model*, as recommend by VITAL [S2], IoT-ARM [S8], RERUM [S32], and IoTEP [S63] shows the semantic and ontological overlay of an IoT based environment, e.g. entities forming the platform. It represents real or virtual objects, software components, and their relationships in an IoT-based solution. Figure 7 shows an example of a domain model representing energy sensors to be deployed in target a to collect data.

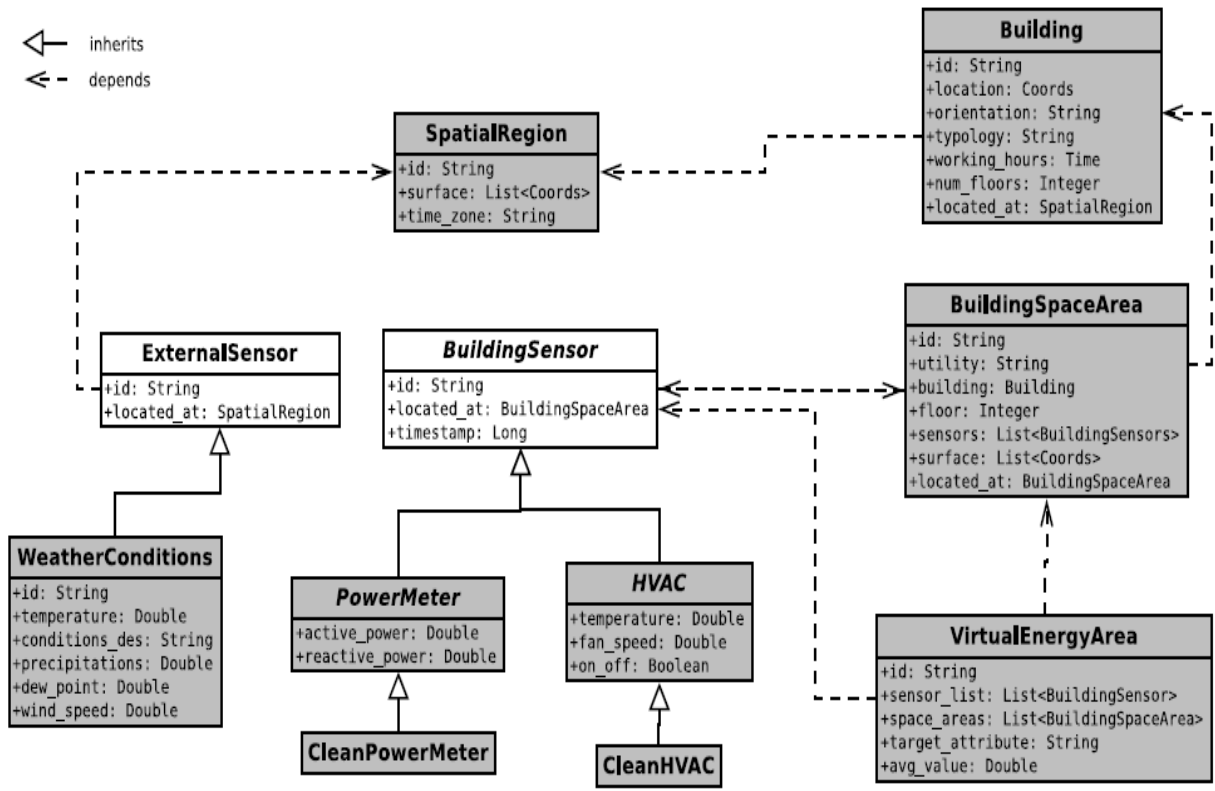

Figure 7. Example of a *domain model* of an energy aware building— source IoTEP [S63]

In traditional software engineering, an architectural view/model of a system helps better understanding of a software system including its structural or behavioural relationships among its components. For example, an architecture models of IoT platform is one the main modelling work-products showing the interacting software and physical components.

Inspired by the 4+1 view model of a system architecture (Kruchten, 1995), IoT-ARM [S8] defines the following models: *domain model*, *requirement model*, *communication model*, *deployment model*, *data model*, *data flow*, *functional model*, *channel model*, *event model*, *context model*, and *road map*. We found that, compared to the all the existing platforms, IoT-ARM includes the most comprehensive collection of prescribed models. The architectural views in IoT-ARM [S8], collectively, show relationships, dependencies, and interactions between platform's components and interfaces to external components, outside world, required smart objects and where they are installed, their relationships e.g. directly or remote, and what physical objects are monitored by sensors. Such similar models are suggested in other approaches. One of the commonly recommended models in the approaches VITAL [S2], IoT-ARM [S8], EADIC [S17], Noise mapping [S27], RERUM [S32], EPIC [S34], SOFIA [S38], OGC [S41], and BET [S60] is a *deployment model*. As name implies, it represents the topology of platform software components and their connection on the physical layer of the platform. Another work-product is the *data follow model*, defined by VITAL [S2], IoT-ARM [S8], FIWARE [S12], Vilajosana [S13], Giang [S21], RERUM [S32], BASIS [S44], SORASC [S46], and IoTEP [S63]. This model represents how data processing is coordinated among platform components or how data entered to smart objects, processed, and then sent out to other smart objects. In addition, the *IoT communication model* is recommended by VITAL [S2], IoT-ARM [S8], FIWARE [S12], Giang [S21], RERUM [S32], and SOFIA [S38]. This model shows how the complexity of communication in heterogeneous IoT environments are handled. Furthermore, an architectural point of view to an IoT platform is the *event model*, which represents how events are triggered and processed by a platform. For example, users interact with the platform by triggering events in devices and services. A formal representation of events should be generated to capture what/who/where/why/when/how events occur to identify necessary functions in the platforms. Developers may model domain entities of the platform. It should be noted that generating models (Table 7) do not happen in a vacuum; instead, a proper understanding of the platform development requirements, its domain, and developers' opinion determine the necessity of generating these models during development process.

Table 7. models prescribed during platform development (Note: - √: addressed, ×:not-addressed)

| Platform | Analysis phase | | | | | Design phase | | | | | | | |
|---|---|---|---|---|---|---|---|---|---|---|---|---|---|
| | City model | Road map/city visions | Requirement model | IoT Canvas | Domain model | Communication model | Deployment model | Data model | Data flow | Functional model | Channel model | Event model | Context model |
| VITAL [S2] | √ | × | √ | × | √ | √ | √ | √ | √ | √ | √ | √ | × |
| CiDAP [S3] | √ | × | × | × | × | × | × | × | × | × | × | × | × |
| IoT-ARM [S8] | × | √ | √ | × | √ | √ | √ | √ | √ | √ | √ | × | √ |
| FIWARE [S12] | × | × | × | × | × | √ | × | × | √ | × | × | × | × |
| Vilajosana[S13] | × | × | × | × | × | × | × | × | √ | × | × | × | × |
| Scallop4SC [S14] | × | × | × | × | × | × | × | √ | × | × | × | × | × |
| EADIC [S17] | × | × | × | × | × | × | √ | × | × | × | × | × | × |
| Giang [S21] | × | × | × | × | × | √ | × | × | √ | × | × | × | × |
| Noise mapping [S27] | × | × | √ | × | × | × | √ | × | × | × | × | × | × |
| RERUM [S32] | × | × | √ | × | × | √ | √ | √ | √ | √ | × | × | × |
| RAMI [S33] | × | × | × | × | × | × | × | √ | × | √ | × | × | × |
| EPIC [S34] | √ | × | × | × | × | × | √ | × | × | × | × | × | × |
| EdSC [S37] | × | × | × | × | × | × | × | × | × | × | × | √ | × |
| SOFIA [S38] | × | × | × | × | × | √ | √ | × | × | × | × | √ | × |
| CityPulse [S39] | × | × | × | × | × | × | × | √ | × | × | × | √ | × |
| SCCM [S40] | × | × | × | × | × | × | × | √ | × | × | × | √ | × |
| OGC [S41] | √ | × | × | × | × | × | √ | √ | × | × | × | × | × |
| BASIS [S44] | × | × | × | × | × | × | × | × | √ | × | × | × | × |
| BSI [S45] | √ | √ | × | × | × | × | × | × | × | × | × | × | × |
| SORASC [S46] | × | × | × | × | × | × | × | × | √ | × | × | √ | × |
| ESPRESSO [S48] | √ | × | × | × | × | √ | √ | √ | × | × | × | √ | × |
| Colin's platform [S55] | × | × | × | × | × | × | × | × | × | × | × | × | × |
| Ignite [56] | × | × | × | × | √ | × | × | × | × | × | × | × | × |
| BET [60] | × | × | × | × | × | × | √ | × | × | × | × | × | × |
| IoTEP [S63] | × | × | × | × | √ | × | × | × | √ | × | × | × | √ |

## 4.6.2 Modelling language

Using a modelling language in the development process of an IoT platform has dual purposes: (i) to represent models during the development time of the platform and (ii) to enable users or platform components to interact at the run-time. These are elaborated in the following.

Firstly, a particular notation and semantic rules are needed to represent models/work-products generated during the platform development activities. A modelling language enables developers to precisely model the different aspects of the target IoT platform. Moreover, using a modelling language is useful to keep the consistency of communication among all stakeholders. The approaches vary in using modelling languages from simple block diagrams to semi-formal languages such as Unified modelling language (UML) (UML, 2004). UML is commonly used in object-oriented software development as suggested by IoT-ARM [S8], EADIC [S17], RERUM [S32], Ignite [S56], and IoTEP [S63]. For example, Ignite [S56] uses the concepts like *use case* and *deployment diagrams* from UML to represent requirement models and the logical deployment of platform components, respectively. RERUM [S32] also suggests using entity-relationship diagram (ERD) to represent data elements, i.e. entities, of the smart city environment as well as their attributes and interrelationships. Our review reveals that other approaches are silent about using or suggesting a modelling language. We discuss this issue in Section 5.4.

Secondly, a modelling language increases the automation and productivity of new IoT application development. In this regards, Salihbegovic et. al. call for designing domain-specific languages (DSML) pertinent to IoT platforms to simplify building new IoT applications (Salihbegovic, Eterovic, Kaljic, & Ribic, 2015). Node-RED, proposed by IBM [S47], is a language supported with a tool for the visually wiring and configuring of IoT devices, APIs, and services together. The language provides a wide range of sensory node elements and data flow patterns. Created models using Node-RED language can be deployed and executed by the platform's engine. VITAL [S2] and IBM [47] are two example platforms using Node-RED. VITAL [S2] offers Node-RED for a rapid development of new back-end IoT services over VITAL [S2].

Recall from section 4.4.3.2, the data processing, which is a key function offered by an IoT platform, can be conducted via a modelling language. This is why SPARQL (SPARQL Protocol and RDF Query Language) is used by SPITFIRE [S20], GAMBAS [S25], and MLSC [S29] for performing queries over the data collected and stored in sensors. A further extension of SPARQL, called Big Data Processing Language (BDPL), is suggested by PLAY [S42] which is used for event processing described in RDF format.

As mentioned in Section 4.4.3.2, a query processor function is to perform queries over data stored in smart objects. Query processors execute queries over data sources. Existing platforms offer different query languages in support of this function. The query language, which is used in the platforms SPITFIRE [S20], GAMBAS [S25], and MLSC [S29], to perform queries over sensors is SPARQL. SPARQL assumes that sensors are represented in a resource discovery format (RDF) triples e.g. sensor type, location, or accuracy. As such, SPARQL enables the query and retrieval of data in the same format. Table 8 shows the list of modelling languages used in the existing approaches.

Table 8 Modelling languages adopted by existing approaches to represent produced models (Note: - √: addressed, ×:not-addressed)

| Platform | development time | | | run time | | | | |
|---|---|---|---|---|---|---|---|---|
| | UML | ERD | Simple Block Diagram | Node-RED | SPARQL | DOLCE | CityGML | BDPL |
| IoT-ARM [S8] | √ | × | × | × | × | × | × | × |
| EADIC [S17] | √ | × | × | × | × | × | × | × |
| RERUM [S32] | √ | √ | √ | × | × | × | × | × |
| VITAL [S2] | × | × | × | √ | × | × | × | × |
| OpenIoT [S9] | × | × | × | × | √ | × | × | × |
| SPITFIRE [S20] | × | × | × | × | √ | × | × | × |
| RERUM [S32] | × | × | × | × | × | √ | × | × |
| OGC [S41] | × | × | × | × | × | × | √ | × |
| PLAY [S42] | × | × | × | × | √ | × | × | √ |
| IBM [S47] | × | × | × | √ | × | × | × | × |
| BET [S60] | √ | × | √ | × | × | × | × | × |
| IoTEP [S63] | √ | × | × | × | × | × | × | × |

# 5 Findings and the future research directions

The review of the selected approaches presented in Section 4 not only provides a comprehensive and fundamental understanding of development processes of IoT platforms, but it also reveals some issues impacting the domain that require further investigation. According to the analysis results of the approaches in Section 4 using our evaluation framework, the following gaps are identified and should be addressed by future IoT platform development approaches.

## 5.1 One-size-fits-all assumption is not a practical choice

It has long been acknowledged that information system development approaches should be customised if they are to achieve optimum effect (Fitzgerald, Hartnett, & Conboy, 2006). In this regard, it is not practical to find or design an IoT development approach that supports all of the features proposed in our evaluation framework. The reason is that an approach may merely focus on some features and not be concerned with other features. Hence, one cannot claim that one approach is superior over another due to missing some features. This implies the fact that the selection of an IoT development approach depends on the requirements and context of an IoT project. For example, if an initial early stage analysis of risks and considerations for IoT developments is required before performing a detailed architecture design and implementation, adopting approaches such as IoT-ARM [S8], SCRM [S19], BSI [S45], and IBM [S47] could be selected as they focuses on initialisation and analysis phases.

On the other hand, developers may wish to address interoperability issues of smart objects at the physical network layer of an IoT architecture. In this regard, the majority of the existing approaches, except for a few such as Vilajosana [S13], Scallop4SC [S14], SCRM [S19], and EPIC [S34] that overlook this feature, would be possible options to accommodate. Similarly, to address the interoperability issue at the software/service layer of an IoT platform, developers are referred to approaches such as VITAL [S2], IoT-ARM [S8], FIWARE [S12], and GAMBAS [S25].

The abovementioned comments signify an important research direction. That is, there is a need to investigate how to design situation-specific IoT platform development approaches which can meet the requirements of a specific project. The need for the situation-specific IoT platform development is highlighted by some of the reviewed approaches SCRM [S19] and BSI [S45]. As a first promising attempt, Girary et.al. (Giray & Tekinerdogan, 2018) suggest a method engineering approach (Harmsen, Brinkkemper, & Oei, 1994) to design customised IoT platform development approach but what key method fragments required and how they should be composed to create a bespoke platform development approach have not yet been addressed in the literature.

## 5.2 Requirements analysis is missing

The initialisation and analysis phases of the IoT platform development in the existing approaches are at infancy stage. Developers may arguably suggest using traditional requirements engineering (RE) techniques available in software engineering literature. This can be plausible to some extent, as such RE techniques are used in the reviewed approaches (e.g. IoT ARM [S8]). In contrast, Alqassem (Alqassem, 2014) argues that an IoT development may face both technical and non-technical issues such as the increasing number of stakeholders at the scale of a city with diverse requirements, responsibility distribution, potential legal issues, and the dynamicity of the smart city environment . An important issue in relation to the requirements analysis is the identification of key stakeholders because IoT projects are typically large in size and involve a variety of stakeholder groups with different objectives, requirements, and commitment levels. Our review reveals that the existing approaches neglect the role of stakeholders. Among the reviewed approaches, we found that only IoT-ARM [S8], SCRM [S19], BSI [S45], and IBM [S47] provide a partial support for the analysis phases to get a wider stakeholder acceptance. It is important to capture stakeholder requirements in a systematic way and to source them to the design phase of the development process. From the above discussion, another research direction is to extend existing approaches or to design IoT-specific requirement analysis techniques that pays attention to stakeholders identification and engagement.

## 5.3 Definition of roles not exploited

Although the approaches define some roles attuned to IoT platform development process, their definitions and responsibilities are not adequately explicated. We believe more research is required for the identification of roles that are specific to IoT context. We suggest a role-driven approach for the IoT development is yet another potential future research. Such an approach has positive contributions to IoT development such as (i) making clear the responsibilities of each stakeholder in the course of an IoT platform development process and maintenance (ii) defining the priority for responsibilities, (iii) specifying necessary interaction and cooperation between the roles, and (iv) enabling more effective IoT project management.

## 5.4 Modelling traceability not explicated

As pointed out in Section 4.6.1, producing a chain of related models plays a key role in an IoT development process. As it can be seen in Table 7, the existing approaches defined a variety of models to be generated as a means to represent different aspects of an IoT architecture. Interestingly, the most recommended models in that table are related to the deployment model and data model. An observable issue is a clear lack of traceability between models during development phases . In other words, there is no clue about how different models are defined and transformed to each other to build the platform. Developers may find different suggested models in the existing approaches but they may need to know how such models are sequentially connected or combined together in the course of the development. A potential future work is based on the fact that since the ultimate goals of a systematic approach is to produce a quality IoT platform, a model-driven approach can be helpful in describing the traceability and seamless transformation of intermediate models that are necessary to reach such a final product.

## 5.5 Testing is not sufficiently supported

An observation from the reviewed approaches is that only Collins [S55] and INTER-METH [S58] provide support for testing activities. This indicates a clear lack of support in the existing approaches when developers need to know techniques to test IoT platforms. A further research direction is to design testing techniques specific to IoT platform development that can be individually adopted or incorporated into the existing approaches.

# 6 Limitations

There are some limitations in the survey presented in this paper in terms of internal and external validity threats. *Internal validity* is related to factors that a researcher has not been aware of and may have affected the research outcome i.e. the evaluation framework and reported analysis results (Wohlin et al., 2012). *External validity* threats are to check the extent to which research outcomes can be generalised (Wohlin et al., 2012).

In terms of internal validity, two issues are notable. Firstly, we focused on analysing works aiming at proposing approaches to build IoT platforms, which yielded in 63 identified papers. Like any other literature review surveys, we cannot guarantee that we have covered all works related to the research questions. The reason is that in an emerging domain like IoT, the literature is full of variety of concepts and viewpoints which are not necessarily consistent. We found that it is hardly any two papers that use the same definition of IoT platform development. An example of this issue is the possible number of layers of IoT architecture that are defined by different papers (e.g. a seven-layer architecture in TSB [S36] and a three-based layer in EdSC [S37]). To alleviate missing any papers, our literature review was not conducted in a linear and mechanical fashion. Instead, we initially started with reading some review papers to get immersed to the field and organise our review. Our literature review was itself a process of understanding the IoT platform development in the sense that we adopted literature review in the early critical reading of the literature, and fine-tuning search strings, inclusion and exclusion criteria, and conducting the review as an understanding of the IoT domain.

Secondly, the reliability of analysis results (Section 4) may have been subjected to the accuracy of the written documents of these approaches. We frequently found that the research conducted in the existing studies including validation techniques, contextual information, and data analysis had not been properly reported. This may have weakened the internal validity of our reported results. As a countermeasure, we tried to find any supplementary documents related to each approach, if required, to ensure the quality of the data extraction.

In terms of external validity, we do not claim that the proposed evaluation framework is complete to include an exhaustive analysis of all approaches to develop IoT platforms. There might be some important features that should have been considered when assessing approaches. At this stage, there is no assertion regarding the generalisability of the evaluation framework beyond 63 identified approaches in this survey. The iterative and gradual refinement of the framework was a technique to minimise any possible feature omission. However, the framework can be extended with new features if it is used to appraise more upcoming IoT platform development approaches that will be introduced in future.

# 7 Related surveys

To the best of our knowledge, no previous survey has undertaken to explore the aspect of engineering lifecycle process of IoT platforms. In the following, we compare and contrast our work in this article and the most related published surveys. This comparison is summarised in Table 9.

We discarded surveys aiming at demystifying the notion of the IoT-based smart cities and discussing prevailing challenges in embarking IoT as they do not narrow down into the aspect of platform development process. Some of the example surveys in this group were trends in the IoT development (Cocchia, 2014), conceptualising and terminology analysis (Macadar, Porto, & Luciano, 2016), sustainable IoT smart city architecture (Trindade et al., 2017), and the governing IoT smart city (Meijer & Bolívar, 2016). Despite their usefulness to get an initial draft of our proposed framework, this class of surveys falls outside the scope and focus of our survey.

Al-Fuqaha et al. give an overview of technical details related to IoT enabling technologies, protocols, and applications to give a view of different communication protocols between heterogeneous existing things such as vehicles, phones, appliances, and goods (Al-Fuqaha et al., 2015). Moreover, we identified the work by Kyriazopoulou who presented a few perspectives in an IoT architecture implementation namely layer-based, service-oriented, event-driven, IoT, and combined architectures from which a set of basic architectural functional and quality factors related to the architecture design are suggested (Kyriazopoulou, 2015). Talari et al. provide a broad overview of applications, practical evidence, and challenges, in particular, security and heterogeneity of smart cities (Talari et al., 2017). The work presented by da Silva et al. highlights the main quality factors such as mobility, sustainability, availability, privacy, and flexibility/extensibility to be fulfilled when implementing an IoT platform (da Silva et al., 2013). Asghari et. al. presents a review and classification of IoT applications including their domain, context, and evaluation factors (Asghari, Rahmani, & Javadi, 2019).

Perhaps, the closest surveys to the proposed dimensions in our framework, but less extensive and centred to the IoT development process, are by (Santana, Chaves, Gerosa, Kon, & Milojicic, 2017) and (Raaijen & Daneva, 2017). Santana et al. have presented a classification of 23 smart city platforms regarding the features such as most enabler technologies (e.g. cyber-physical systems, IoT, big data, and cloud computing). They also compare a selected set of exiting platforms with respect to functional requirements and quality factors that are expected to be addressed by an IoT platform. Similarly, Raaijen et. al. (Raaijen & Daneva, 2017) has identified a set of technical and non-technical challenges in an IoT development. They have reviewed 29 studies and conducted a field study through which a framework of influential factors is derived which is useful for policy-makers in order to assess the viability of an IoT architecture. Whilst our framework has been inspired by the ideas presented in the aforementioned surveys, we extended them with new important characteristics in the following ways:

— *Focus and depth of analysis*. Our survey supersedes the existing ones by adding a new perspective to the extant material in the literature as it concentrates on the development process of IoT platforms. It limits its view to the existing proposals providing either a complete or a partial approach for the development of IoT platforms. Thus, it is more to the point compared to the related surveys. The proposed evaluation framework encompasses four different aspects (Figure 1), which have not been covered by any of the existing surveys. In addition, the existing surveys do not provide a comprehensive discussion of the quality factors related to the architectural design phase. For example, the features presented in the survey by (Santana et al., 2017) cover no more than 7 features related to the *design phase* in our framework. Another distinct feature of the current survey is to provide a deeper explanation of the quality factors related to the design phase.

— *Survey coverage*. Due to our comprehensive analytical lens to critique the literature, we have covered different and more recent published approaches that are missing by other surveys. For example, we found that only 5 out of our 63 reviewed works in the current survey have been evaluated by (Santana et al., 2017). Therefore, this survey is a complementary to the related surveys.

Table 9. Comparison of our survey and the related surveys

| Survey | Focus and depth of analysis | Number of papers |
|---|---|---|

| (Santana et al., 2017) | Functional and quality factors of IoT platforms | 23 |
|---|---|---|
| (Raaijen & Daneva, 2017) | Technical and non-technical challenges of IoT smart city development | 29 |
| (Al-Fuqaha et al., 2015) | Enabling technologies, protocols, and applications of IoT platforms | 14 |
| (Kyriazopoulou, 2015) | Analysing six architectural perspectives, e.g. SOA and layering, to IoT design | 32 |
| (Talari et al., 2017) | Technologies, barriers to implementation, and applications of IoT platforms | Not stated |
| (da Silva et al., 2013) | Quality factors of IoT platforms | 18 |
| (Asghari et al., 2019) | IoT application domains | 72 |
| Our survey | IoT platform development process | 63 |

# 8 Conclusion

IoT platforms are complex and multifaceted IT artefacts. Using systematic approaches are acclaimed to aid developers to manage the complexity of development and maintenance of IoT platforms in more cohesive and disciplined manner. An ad-hoc approach may result in poor and costly platform maintenance. In this spirit, we presented a detailed review of 63 extant approaches for the development of IoT platforms and highlighted their key characteristics in the view of our proposed framework. A specific advantage of our framework is its usefulness as a tool to select approaches as to satisfy specific requirements of a given IoT platform development scenario.

Our review also revealed important gaps in the existing literature that call for further investigations. Firstly, few works exist on providing a foundation for situation-specific IoT platform development. The rationale for this argument is that each smart city projects may entail different requirements such as city culture, heterogeneity of smart objects, and geographic distribution. To fill this gap, we suggested employing a situational method engineering approach as a research agenda for designing bespoke IoT architecture development approaches. In addition, the current survey also calls for developing new IoT specific requirement engineering techniques that can address the complexity of large scale IoT architectures as early as possible in a platform development endeavour. Furthermore, we found a lack of clarification of roles that participate in IoT development activities. Another identified area for more exploration is that the existing approaches suffer from defining a chain of model traceability and transformation. IoT developers may need to know the key behavioural and structural models that should be generated and mapping together toward developing a target platform. We suggested adopting model-driven approach to alleviate this issue.

As the final note, our survey provides a solid content of important features, recommendations, mechanism, and quality factors that are commonly incorporated into the development process of IoT platforms in one place. Such a comprehensive inventory, which sheds light into the essence of IoT platform development process and can be used by both platform providers and academia, is a significant contribution of this survey. Our second contribution is the proposed evaluation framework. It supports well informed decision making on the analysing and selecting of IoT platform development approaches. Finally, another important utility of our work is that this survey is helpful for novice practitioners and researchers who are interested in understanding how an IoT platform should be developed.

# Appendix A

List of the reviewed approaches

| ID | Authors and title | Acronym | Channel | Source | Year |
|---|---|---|---|---|---|
| [S1] | Eduardo Santana, Zambom Felipe, et al., *"Software platforms for smart cities: Concepts, requirements, challenges, and a unified reference architecture"* | RASCP | Journal | ACM Computing Surveys | 2017 |
| [S2] | Riccardo Petrolo, Valeria Loscri, et al., *"Towards a Smart City based on Cloud of Things, a survey on the smart city vision and paradigms"* | VITAL | Conference | IEEE | 2014 |
| [S3] | Bin Cheng, Salvatore Longo, et al. , *"Building a Big Data Platform for Smart Cities: Experience and Lessons from Santander"* | CiDAP | International Congress | IEEE | 2015 |
| [S4] | Zaheer Khan, Ashiq Anjum, et al., *"Cloud Based Big Data Analytics for Smart Future Cities"* | - | Conference | IEEE/ACM | 2013 |
| [S5] | Jayavardhana Gubbi, Rajkumar Buyya, et al., *"Internet of Things (IoT): A Vision, Architectural Elements, and Future Directions"* | - | Journal | Elsevier | 2013 |
| [S6] | Cisco, *"The Internet of Things Reference Model"* | Cisco | White paper | Cisco | 2014 |
| [S7] | Federico Ciccozzi, Crnkovic Ivica, *"Model-driven engineering for mission-critical IoT systems"* | MC-IoT | Journal | IEEE | 2017 |
| [S8] | Sebastian Lange, Andreas Nettsträter, et al., *"Introduction to the architectural reference model for the Internet of Things"* | IoT-ARM | White paper | IoT-A | 2013 |
| [S9] | John Soldatos, Nikos Kefalakis, *"OpenIoT: Open source Internet-of-Things in the cloud"* | OpenIoT | Workshop | Springer | 2015 |
| [S10] | Jasmin Guth, Uwe Breitenbucher, et al., *"Comparison of IoT Platform Architectures: A Field Study based on a Reference Architecture"* | - | Conference | IEEE | 2016 |
| [S11] | Ivan Ganchev, Zhanlin Ji, et al., *"A Generic IoT Architecture for Smart Cities"* | - | Conference | IEEE | 2014 |
| [S12] | *"FIWARE (also called Open & Agile Smart Cities (OASC))"* | FIWARE/OASC | Technical report | FIWARE Community | 2014 |
| [S13] | Ignasi Vilajosana, Jordi Llosa, et al., *"Bootstrapping smart cities through a self-sustainable model based on big data flows"* | - | Magazine | IEEE | 2013 |
| [S14] | Kohei Takahashi, Shintaro Yamamoto, et al., *"Design and implementation of service API for large-scale house log in smart city cloud"* | Scallop4SC | Conference | IEEE | 2012 |
| [S15] | Zaheer Khan, Ashiq Anjum, et al., *"Towards cloud based big data analytics for smart future cities"* | - | Journal | Springer Open Journal | 2015 |
| [S16] | Catherine E. A. Mulligan, Magnus Olsson, *"Architectural Implications of Smart City Business Models: An Evolutionary Perspective"* | - | Magazine | IEEE | 2013 |
| [S17] | George Kakarontzas, Leonidas Anthopoulos, et al. *"A Conceptual Enterprise Architecture Framework for Smart Cities, A Survey Based Approach"* | EADIC | Conference | IEEE | 2014 |
| [S18] | David Díaz Pardo de Vera, Álvaro Sigüenza Izquierdo, et al. *"A Ubiquitous sensor network platform for integrating smart devices into the semantic sensor web"* | Telco USN-Platform | Journal | Sensors | 2014 |
| [S19] | Sotiris Zygiaris, *"Smart City Reference Model: Assisting Planners to Conceptualize* | SCRM | Journal | Springer | 2012 |
| | *the Building of Smart City Innovation Ecosystems"* | | | | |
| [S20] | Dennis Pfisterer, Kay Romer, *"SPITFIRE: Towards a Semantic Web of Things"* | SPITFIRE | Magazine | IEEE | 2011 |
| [S21] | Nam K Giang, Rodger Lea, et al., *"On Building Smart City IoT Applications: a Coordination-based Perspective"* | - | Workshop | ACM | 2016 |
| [S22] | Rong Wenge, Xiong Zhang, et al., *"Smart City Architecture: A Technology Guide for Implementation and Design Challenges"* | - | Journal | IEEE | 2014 |
| [S23] | Paul Fremantle, *"A reference architecture for the internet of things"* | WSO2 | Technical report | WSO2 | 2015 |
| [S24] | Andrea Zanella, Senior Member, *"Internet of Things for Smart Cities"* | Padova | Journal | IEEE | 2014 |
| [S25] | Wolfgang Apolinarski, Umer Iqbal, et. al., *"The GAMBAS Middleware and SDK for Smart City Applications"* | GAMBAS | Workshop | IEEE | 2014 |
| [S26] | Nader Mohamed, Jameela Al-Jardoodi, *"SmartCityWare: A Service-Oriented Middleware for Cloud and Fog Enabled Smart City Services"* | SmartCityWare | Conference | IEEE | 2017 |
| [S27] | Jiong Jin, Jayavardhana Gubbi, *"An information framework for creating a smart city through internet of things"* | Noise mapping | Conference | IEEE | 2013 |
| [S28] | Panagiotis Vlacheas, Vera Stavroulaki, et al., *"Enabling Smart Cities through a Cognitive Management Framework for the Internet of Things"* | - | Magazine | IEEE | 2013 |
| [S29] | Aditya Gaura, Bryan Scotneya, et. al., *"Smart City Architecture and its Applications based on IoT"* | MLSC | Symposium | Elsevier | 2015 |
| [S30] | Zhihong Yang, Yufeng Peng, et al., *"Study and Application on the Architecture and Key Technologies for IOT"* | - | Conference | IEEE | 2011 |
| [S31] | Miao Wu, Ting-lie Lu, et al. *"Research on the architecture of Internet of things"* | TMN | Conference | IEEE | 2010 |
| [S32] | Henrich C. Pohls, Vangelis Angelakis, *"RERUM: Building a Reliable IoT upon Privacy- and Security- enabled Smart Objects"* | RERUM | Workshop | IEEE | 2014 |
| [S33] | ZAEI, *"Reference Architecture Model Industry 4.0 (RAMI 4.0)"* | RAMI | White Paper | ZAEI | 2015 |
| [S34] | Pieter Ballon, Julia Glidden, *"EPIC Platform and Technology Solution"* | EPIC | White Paper | EPIC | 2013 |
| [S35] | Kenji Tei, Levent Gürgen, *"ClouT : Cloud of Things for Empowering the Citizen Clout in Smart Cities"* | ClouT | Conference | IEEE | 2014 |
| [S36] | Arup, *"Solutions for Cities: An analysis of the feasibility studies from the Future Cities Demonstrator Programme"* | TSB | White Paper | Smart City Strategies A Global Review - ARUP | 2013 |
| [S37] | Liviu-Gabriel Cretu, Alexandru Ioan, *"Smart Cities Design using Event-driven Paradigm and Semantic Web"* | EdSC | Journal | Inforec Association | 2012 |
| [S38] | Luca Filipponi, Andrea Vitaletti, *"Smart City: An Event Driven Architecture for Monitoring Public Spaces with Heterogeneous Sensors"* | SOFIA | Conference | IEEE | 2010 |
| [S39] | Dan Puiu, Payam Barnaghi, et. al., *"CityPulse: Large Scale Data Analytics Framework for Smart Cities"* | CityPulse | Journal | IEEE | 2016 |
| [S40] | ISO, *"ISO/IEC 30182: Smart city concept model — Guidance for establishing a model for data interoperability"* | SCCM | Technical report | ISO (International Organization for Standardization) | 2017 |

| [S41] | Open Geospatial Consortium, *"OGC Smart Cities Spatial Information Framework"* | OGC | White paper | Open Geospatial Consortium | 2015 |
|---|---|---|---|---|---|
| [S42] | Roland Stühmer, Yiannis Verginadis, *"PLAY: Semantics-Based Event Marketplace"* | PLAY | Conference | Springer | 2013 |
| [S43] | M. Nitti, *"IoT Architecture for a Sustainable Tourism Application in a Smart City Environment"* | - | Journal | Hindawi | 2017 |
| [S44] | Carlos Costa, Maribel Yasmina Santos, *"BASIS: A Big Data Architecture for Smart Cities"* | BASIS | Conference | IEEE | 2016 |
| [S45] | BSI, *"Smart city framework – Guide to establishing strategies for smart cities and communities"* | BSI | Technical report | British Standards Institution | 2014 |
| [S46] | S. J. Clement, D. W. McKee, *"Service-Oriented Reference Architecture for Smart Cities"* | SORASC | Conference | IEEE | 2017 |
| [S47] | Arundhati Bhowmick, Eduardo Francellino, et. al., *"IBM Intelligent Operations Center for Smarter Cities Administration Guide"* | - | Book | IBM | 2012 |
| [S48] | Andy Cox, Peter Parslow, et. al., *"ESPRESSO (systEmic Standardisation apPRoach to Empower Smart citieS and cOmmunities)"* | ESPRESSO | White paper | ESPRESSO community | 2016 |
| [S49] | Arthur de M. Del Esposte , Fabio Kon, *"InterSCity: A Scalable Microservice-based Open Source Platform for Smart Cities"* | InterSCity | Conference | Scitepress digital library | 2017 |
| [S50] | Raffaele Giaffreda, *"iCore: a cognitive management framework for the internet of things"* | iCore | Conference | Springer | 2013 |
| [S51] | Andreas Kamilaris, Feng Gao, *"Agri-IoT: A Semantic Framework for Internet of Things-enabled Smart Farming Applications"* | Agri-IoT | Conference | IEEE | 2016 |
| [S52] | Yong Woo Lee, Seungwoo Rho, *"U-City Portal for Smart Ubiquitous Middleware"* | U-City | Conference | IEEE | 2010 |
| [S53] | Chayan Sarkar,Akshay Uttama Nambi S. N., *"DIAT: A Scalable Distributed Architecture for IoT"* | DIAT | Journal | IEEE | 2015 |
| [S54] | Gilles Privat, et. al. *"Towards a Shared Software Infrastructure for Smart Homes, Smart Buildings and Smart Cities"* | SmartSantander | Workshop | - | 2014 |
| [S55] | Tom Collins, *"A Methodology for Building the IoT"* | Collins | - | - | 2014 |
| [S56] | Frank Puhlmann, Dirk Slama, *"An IoT Solution Methodology"* | Ignite | - | - | Not stated |
| [S57] | C. Savaglio, *"A Methodology for the Development of Autonomic and Cognitive Internet of Things Ecosystems"* | ACOSO-Meth | Thesis | - | 2017 |
| [S58] | G. Fortino, R. Gravina, et. al., *"A Methodology for Integrating Internet of Things Platforms"* | INTER-METH | Conference | IEEE | 2018 |
| [S59] | Marcello A. Gómez Maureira, Daan Oldenhof, et al., *"ThingSpeak–an API and Web Service for the Internet of Things"* | ThingSpeak | Conference | World Wide Web | 2011 |
| [S60] | Venticinque Salvatore, Alba Amato, *"A methodology for deployment of IoT application in fog"* | BET | Journal | Springer | 2019 |
| [S61] | Amany Sarhan, *"Cloud-based IoT Platform: Challenges and Applied Solutions"* | Galliot | Journal | IGI Global | 2019 |

| [S62] | Alvaro Luis Bustamante , Miguel A. Patricio, "*Thinger.io: An Open Source Platform for Deploying Data Fusion Applications in IoT Environments*" | Thinger.io | Journal | Sensor | 2019 |
|---|---|---|---|---|---|
| [S63] | Fernando Terroso-Saenz, Aurora González, et. al. , "*An open IoT platform for the management and analysis of energy data*" | IoTEP | Journal | Elsevier | 2019 |

# Appendix B

Research quality criteria adapted from (Greenhalgh & Taylor, 1997) and (Kitchenham et al., 2002)

| Criteria | Type | Evaluation question(s) | Possible values |
|---|---|---|---|
| Research aim | Scale | (i) Is there a clear statement of research objective?<br>(ii) Is there an explanation of research rationale? | ● Completely described.<br>◑ Considerably described.<br>◒ Moderately described.<br>◔ Slightly described.<br>○ Not at all. |
| Research design | Scale | (i) Is there any description of the research techniques to conduct the research?<br>(ii) Is there any description of type of the research in terms of being empirical, qualitative, or design science, or mixed method, etc.? | ● Completely described.<br>◑ Considerably described.<br>◒ Moderately described.<br>◔ Slightly described.<br>○ Not at all. |
| Data collection | Scale | (i) Is there any clear description of data collection method (e.g. recruitment strategy) appropriate to the research?<br>(ii) Is there any description of measurement used to collect data?<br>(iii) Is there a deception of mechanism to collect data e.g. interview, survey, domain expert review?<br>(iv) Is there a way through which data are recorded e.g. tape recording, video material, note etc.? | ● Completely described.<br>◑ Considerably described.<br>◒ Moderately described.<br>◔ Slightly described.<br>○ Not at all. |
| Data analysis | Scale | (i) Is there a clear and rigors description of data analysis?<br>(ii) Is there a clear description of the tool used to analysis data? If data analysis was performed manual, is there a clear description of the way used for data analysis? | ● Completely described.<br>◑ Considerably described.<br>◒ Moderately described.<br>◔ Slightly described.<br>○ Not at all. |
| Reflexivity | Scale | (i) Is there an appropriate description of the relationship between researcher and participants defined to conduct research?<br>(ii) Is there any clarification of potential bias and influence of researcher or external factors on data collection, analysis, and report? | ● Completely described.<br>◑ Considerably described.<br>◒ Moderately described. |
| Value | Scale | (i) Is there a clear description of contributions to research and practice?<br>(ii) Is there a clear description of research justification and significance? | ◑ Slightly described.<br>○ Not at all.<br><br>● Completely described.<br>◕ Considerably described.<br>◓ Moderately described.<br>◔ Slightly described.<br>○ Not at all. |
| Findings | Scale | (i) Is there a clear description of research findings/outcomes?<br>(ii) Is there any justification, discussion, or evidence of research findings? | ● Completely described.<br>◕ Considerably described.<br>◓ Moderately described.<br>◔ Slightly described.<br>○ Not at all. |
| Validation | Multiple | Has the architecture been validated in real world scenario? | Case study, exemplar scenario, domain expert review, survey, and simulation, no validation |

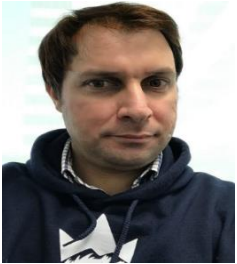

**Dr Mahdi Fahmideh** is lecturer (assistant professor) of Information Technology in the Faculty of Engineering and Information Sciences at University of Wollongong, Australia. He received a PhD degree in Information Systems from the Business School of University of New South Wales, Sydney, Australia. His research focuses on creating new-to-the-world artifacts that help organizations in adopting IT initiatives to tackle problems. His research outcome can be in the form of methodological approaches, conceptual models, decision-making frameworks, and software tools. His research interests lie at the intersection of cloud computing, data analytics, IoT, blockchain, and method engineering. Prior to joining the academia, Dr. Fahmideh has worked as a software developer in different industry sectors including accounting, insurance, defence, and publishing.

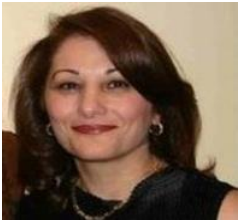

**Dr Didar Zowghi** is professor of Software Engineering in the Faculty of Engineering and Information Technology at the University of Technology Sydney (UTS), Australia. Her research focuses on improving software development processes and the quality of their products. She has completed many research projects in requirements engineering, global software development, technology adoption, web technologies, software process improvement, service oriented computing, data quality in healthcare, IoT/Smart Cities and mobile learning. Professor Zowghi is Associate Editor of IEEE Software and regional editor of Requirements Engineering journal and IET Software journal. She has co-authored papers with 90 different researchers from 30 countries.